\documentclass[acmsmall,screen,anonymous=false]{acmart}
\usepackage{amsmath,microtype,siunitx,xspace}
\usepackage{hyperref,cleveref,graphicx,syntax}
\usepackage{listings,subcaption,multirow,mathpartir}

\newcommand{\Impl}[2]{\ensuremath{\F{Impl}\!\left\langle #1, #2\right\rangle}}
\newcommand{\str}[1]{\text{``\F{#1}''}}
\newenvironment{code}{\begin{tabular}{>{$}l<{$}}}{\end{tabular}}
\renewcommand{\>}{\quad\quad}
\newcommand{\sci}[2]{#1\text{\textsc{e}}{#2}}

\setcopyright{acmcopyright}
\copyrightyear{2023}
\acmYear{2023}
\acmDOI{XXXXXXX.XXXXXXX}



\newcommand{\todo}[1]{{\color{red}{[#1]}}}
\newcommand{\F}[1]{\text{\textsf{#1}}}

\newcommand{\name}{MegaLibm\xspace}

\begin{document}

\title{Implementation and Synthesis of Math Library Functions}

\author{Ian Briggs}
\email{ibriggs@cs.utah.edu}
\affiliation{%
  \institution{University of Utah}
  \city{Salt Lake City}
  \state{UT}
  \country{USA}
}

\author{Yash Lad}
\email{yash.lad@utah.edu}
\affiliation{%
  \institution{University of Utah}
  \city{Salt Lake City}
  \state{UT}
  \country{USA}
}

\author{Pavel Panchekha}
\email{pavpan@cs.utah.edu}
\affiliation{%
  \institution{University of Utah}
  \city{Salt Lake City}
  \state{UT}
  \country{USA}
}

\begin{abstract}
Achieving speed and accuracy
  for math library functions like \F{exp}, \F{sin}, and \F{log}
  is difficult.
This is because low-level implementation languages like C
  do not help math library developers
  catch mathematical errors,
  build implementations incrementally,
  or separate high-level and low-level decision making.
This ultimately puts development of such functions out of reach
  for all but the most experienced experts.
To address this, we introduce \name, a domain-specific language
  for implementing, testing, and tuning math library implementations.
\name is safe, modular, and tunable.
Implementations in \name can automatically detect
  mathematical mistakes like sign flips
  via semantic wellformedness checks, and
  components like range reductions
  can be implemented in a modular, composable way,
  simplifying implementations.
Once the high-level algorithm is done,
  tuning parameters like working precisions and evaluation schemes
  can be adjusted through orthogonal tuning parameters
  to achieve the desired speed and accuracy.
\name also enables math library developers to work interactively,
  compiling, testing, and tuning their implementations
  and invoking tools like Sollya and type-directed synthesis
  to complete components and synthesize entire implementations.
\name can express 8 state-of-the-art math library implementations
  with comparable speed and accuracy to the original C code,
  and can synthesize 5 variations and 3 from-scratch implementations
  with minimal guidance.
\end{abstract}

\maketitle

\section{Introduction}
\label{sec:introduction}

Mathematical computations in tasks as diverse
  as aeronautics, banking, scientific simulations, and data analysis
  are typically implemented as operations on floating-point numbers.
The basic operators---%
  addition, subtraction, multiplication,
  and possibly division, square roots, and fused multiply-adds---%
  are typically provided by the hardware,
  but higher-level mathematical functions
  such as trigonometric or exponential functions
  are implemented in software in libraries such as \F{libm}.
The speed and accuracy of these software libraries
  can have a dramatic impact
  on applications such as 3D graphics~\cite{optuner}.

To maximize performance,
  math libraries are written
  in low-level languages like C;
  \Cref{fig:fdlibmlog} shows one example.
Ensuring correctness and accuracy is thus challenging.
These implementation languages
  cannot prevent mathematical errors
  such as mixing up signs or using the wrong transformations,
  so the developer must be constantly vigilant.
Moreover,
  these implementation languages require interleaving
  conceptually distinct components of a math library function
  (such as function approximation
  and range reduction and reconstruction),
  so the developer must hold the implementation's complexity
  in their head.
Finally,
  these implementation languages
  demand making low-level decisions
  such as polynomial evaluation scheme or working precision
  up front,
  so the developer must have the skill
  to make these decisions well before even running the code.
The vigilance, expertise, and experience required
  is a high bar for even the most expert programmers.

\begin{figure}[p]
    \begin{flushleft}
      \Small \it
      Copyright (C) 1993-2004 by Sun Microsystems, Inc. All rights reserved.
      Developed at SunSoft, a Sun Microsystems, Inc. business.
      Permission to use, copy, modify, and distribute this
      software is freely granted, provided that this notice
      is preserved.
    \end{flushleft}

    \begin{footnotesize}
\begin{lstlisting}
double __ieee754_log(double x) {
    double hfsq,f,s,z,R,w,t1,t2,dk;
    int k,hx,i,j;
    unsigned lx;
    hx = __HI(x);		/* high word of x */
    lx = __LO(x);		/* low  word of x */
    k=0;
    if (hx < 0x00100000) {			/* x < 2**-1022  */
        if (((hx&0x7fffffff)|lx)==0)
        return -two54/zero;		/* log(+-0)=-inf */
        if (hx<0) return (x-x)/zero;	/* log(-#) = NaN */
        k -= 54; x *= two54; /* subnormal number, scale up x */
        hx = __HI(x);		/* high word of x */
    }
    if (hx >= 0x7ff00000) return x+x;
    k += (hx>>20)-1023;
    hx &= 0x000fffff;
    i = (hx+0x95f64)&0x100000;
    __HI(x) = hx|(i^0x3ff00000);	/* normalize x or x/2 */
    k += (i>>20);
    f = x-1.0;
    if((0x000fffff&(2+hx))<3) {	/* |f| < 2**-20 */
        if(f==zero) if(k==0) return zero;  else {dk=(double)k;
                 return dk*ln2_hi+dk*ln2_lo;}
        R = f*f*(0.5-0.33333333333333333*f);
        if(k==0) return f-R; else {dk=(double)k;
                 return dk*ln2_hi-((R-dk*ln2_lo)-f);}
    }
    s = f/(2.0+f);
    dk = (double)k;
    z = s*s;
    i = hx-0x6147a;
    w = z*z;
    j = 0x6b851-hx;
    t1= w*(Lg2+w*(Lg4+w*Lg6));
    t2= z*(Lg1+w*(Lg3+w*(Lg5+w*Lg7)));
    i |= j;
    R = t2+t1;
    if(i>0) {
        hfsq=0.5*f*f;
        if(k==0) return f-(hfsq-s*(hfsq+R)); else
             return dk*ln2_hi-((hfsq-(s*(hfsq+R)+dk*ln2_lo))-f);
    } else {
        if(k==0) return f-s*(f-R); else
             return dk*ln2_hi-((s*(f-R)-dk*ln2_lo)-f);
    }
}
\end{lstlisting}
\end{footnotesize}

\caption{
  The C source code for  fdlibm's $\log(x)$ function.
  Constant and macro definitions omitted for space.
}
\label{fig:fdlibmlog}
\Description{The source code of fdlibm's $\log(x)$.}

\end{figure}

We propose a new way of implementing math libraries
  that circumvents these problems.
At the core of our approach is a new DSL, \name,
  which allows expressing
  the high-level algorithms behind an implementation
  without fixing low-level tuning decisions;
  the \name implementation of \Cref{fig:fdlibmlog}
  is shown in \Cref{fig:mlmlog}.
The implementation in \name
  can then be compiled to fast and accurate C code,
  and low-level control over the compiled code is available
  through optional tuning parameters
  orthogonal to the high-level algorithm.
Developers can thus
  to make, test, and tweak low-level choices
  interactively, in a Jupyter notebook, based on measurement and data.
Specialized, application-specific math libraries
  can thus be rapidly written, tested, and tuned.

The key to this workflow are three properties of the \name DSL:
  safety, modularity, and tunability.
Safety means that mathematical mistakes,
  such as putting the wrong sign
  or mis-parenthesizing an expression,
  are caught at compile time
  by the type checker
  via a set of semantic wellformedness rules.
Modularity means
  that complex math function implementation components
  such as range reduction and reconstruction
  are expressed via a compositional theory of function identities,
  allowing complex range reductions
  to be expressed as combinations of simple ones.
And tuning means that \name carefully separates
  real-number from floating-point reasoning,
  allowing low-level speed and accuracy decisions to be made
  separately from high-level algorithmic ones.
These properties are ensured by careful design of the \name DSL.
This DSL also permits the integration of automated synthesis tools.
The \name type system, for example,
  provides the information needed to invoke
  function approximation tools such as Sollya,
  while \name's semantic wellformedness rules
  allow for type-directed synthesis
  of range reduction and reconstruction algorithms.
By leveraging the \F{egg} e-graph library~\cite{egg},
  including new features such as
  node extraction and e-graph intersection,
  even synthesis of whole function implementations from scratch
  becomes practical.

We demonstrate that \name can support
  expert, state-of-the-art implementations
  by replicating 8 function implementations
  from the AOCL~\cite{amdlibm}, fdlibm~\cite{fdlibm}, and VDT~\cite{vdt} math libraries,
  achieving comparable accuracy and performance.
We then design 5 variations on these implementations,
  achieving either greater accuracy or performance,
  to demonstrate that \name enables
  experimentation and design space exploration.
Finally, we show that \name can synthesize 3 implementations
  of related functions \F{sinpi}, \F{vercos}, and \F{sinmx}
  with no input from the user.
Section 8 discusses limitations and future work.

\smallskip
\noindent
In short, this paper contributes:
\begin{itemize}
\item A new DSL, \name,
  for safe, modular, and tunable math library implementation
  (\Cref{sec:dsl}).
\item Interactive tooling to rapidly develop math function implementations in \name
  (\Cref{sec:tools}).
\item Automated synthesis of math function implementations with minimal guidance
  (\Cref{sec:synthesis}).
\end{itemize}
\Cref{sec:overview} provides an overview
  of math library implementation and \name,
  \Cref{sec:eval} contains our evaluation of \name,
  and \Cref{sec:related} summarizes the extensive literature
  on math library implementation.

\section{Overview and Background}
\label{sec:overview}

This overview walks the reader
  through implementing $\cos(x)$ in \name
  in the interactive style of a Jupyter notebook session.

\subsection{The \name DSL}

At their core, most math function implementations
  \emph{approximate} the target function
  as some easier-to-evaluate function.
For example, polynomials are easy to evaluate
  because they only use addition and multiplication,
  and by picking the right polynomial,
  almost any function can be approximated with little error.
For example, the venerable Taylor approximation of $\cos(x)$ is
  $1 - \frac{x^2}{2!} + \frac{x^4}{4!} + \ldots$,
  which is a polynomial with coefficients $1$ for $x^0$,
  $-\frac12$ for $x^2$, and $\frac1{24}$ for $x^4$.
Polynomial approximations are also
  the simplest kind of \name implementation,
  expressed using a \F{polynomial} term:%
\footnote{Expert readers, fear not:
  we will replace the Taylor approximation with a better one
  later in this section.}
\begin{align*}
  & a_0 = 1, a_2 = -1/2, a_4 = 1/24 \\
& \F{impl}_1 = \F{polynomial}\left(\{0: a_0, 2: a_2, 4: a_4\}\right)
:
\Impl{a_0 + a_2 x^2 + a_4 x^4}{[-\infty, \infty]}
\end{align*}
Here, \F{polynomial} takes a dictionary
  mapping powers of $x$ to the corresponding coefficient
  and returns a \name term of \F{Impl} type.
Strictly speaking, this term represents
  an \textit{implementation} of a polynomial,
  with the polynomial itself, and its input domain,
  given by the parameters of its \F{Impl} type.
Distinguishing between mathematical functions
  and their implementations is essential in \name,
  because there are typically many non-equivalent ways
  to compute a polynomial in floating-point arithmetic:
  different evaluation schemes,
  different or mixed precisions, and so on.
These subtleties are accessible in \name via tuning parameters,
  which we will discuss later.

Next, we need to inform \name
  that this polynomial is intended to implement $\cos(x)$.
This requires a ``cast'' from an \F{Impl} of a polynomial
  to an \F{Impl} of $\cos(x)$.
This is an easy place to make a mistake or a typo,
  such as mistyping one of the coefficients or dropping a minus sign.
\name's cast operator, \F{approx}, prevents these mistakes
  by requiring an explicit domain and approximation error:
\[
\F{impl}_2 = \F{approx}\left(\cos(x), \left[0, \frac{\pi}2\right], 0.02, \F{impl}_1\right)
:
\Impl{\cos(x)}{\left[0, \frac\pi2\right]}
\]
This allows \name to check
  that the polynomial we have chosen
  is within the given error, $0.02$,
  of the target function, $\cos(x)$.
These \emph{semantic wellformedness} checks
  exist for most terms in \name
  and prevent typos and mathematical mistakes,
  a property we call \emph{safety}.

Besides approximating the target function,
  most math libraries also use
  \emph{range reduction and reconstruction}
  to expand the domain on which an implementation is usable.
These remap inputs from some wider domain,
  like $[-\frac\pi2, \frac\pi2]$,
  to a narrower one like the $[0, \frac\pi2]$ chosen above.
This is necessary because polynomial approximations
  are typically most accurate on a narrow domain.
For \F{cos}, one useful remapping leverages the fact
  that $\cos(-x) = \cos(x)$.
This means that to evaluating \F{cos} on a negative input like $-1$,
  we can instead evaluate \F{cos} on its negation, $1$,
  and still get the right answer.
This technique is is represented in \name via the \F{left} construct,
  which takes three arguments:
  a reduction function (here, negation),
  the implementation to call on the narrower domain
  (here, the implementation from above)
  and a reconstruction function (described later):
\[
\F{impl}_3 = \F{left}(-x, \F{impl}_2, y)
:
\Impl{\cos(x)}{\left[-\frac\pi2, \frac\pi2\right]}
\]
Note that the domain has gone
  from $[0, \frac\pi2]$ to the larger range $[-\frac\pi2, \frac\pi2]$.
In C, this \F{left} term corresponds to the branch \texttt{if (x < 0.0) x = -x}.
Just like \F{approx} terms, range reduction terms
  are checked for semantic wellformedness
  to prevent errors.
For this \F{left} term, the  semantic wellformedness condition
  requires that $s(x) = -x$ maps $[-\frac\pi2, 0]$ to $[0, \frac\pi2]$
  and that $\cos(x) = \cos(-x)$.
If we made a mistake---%
  mistyping the range reduction function,
  or using the wrong input range---%
  the user would see a type error at compile time
  instead of misleading results when running the code.

Real-world range reductions can sometimes require
  \emph{reconstruction} afterwards.
For example, the identity $\cos(x) = -\cos(\pi - x)$
  remaps inputs $x$ to $\pi - x$,
  but only if the output $\cos(\pi - x)$
  is then \emph{reconstructed} by negating it.
The corresponding \name term is:
\[
\F{impl}_4 = \F{right}(\pi-x, \F{impl}_2, -y)
:
\Impl{\cos(x)}{\left[0, \pi\right]}
\]

There are many other cosine identities
  that could be used for range reduction and reconstruction,
  either through \F{left} and \F{right}
  or through \name's other range reduction operators.
There is thus a dizzying array
  of possible range reduction and reconstruction algorithms.
In \name, this design space can be explored modularly
  by nesting range reduction terms.
For example, combining the two reductions we've already seen
  produces an implementation valid from $-\pi$ to $\pi$:
\[
\F{impl}_5 =
\F{left}(-x,
\F{right}(\pi-x, \F{impl}_2, -y),
y)
:
\Impl{\cos(x)}{\left[-\pi, \pi\right]}
\]
This modularity
  allows complex range reduction and reconstruction algorithms
  to be built step-by-step and interactively.
For now,
  let's put more complex range reduction and reconstruction to the side
  and switch to tuning our implementation.

\subsection{Tuning \name Implementations}

\begin{figure}
  \includegraphics[width=\linewidth]{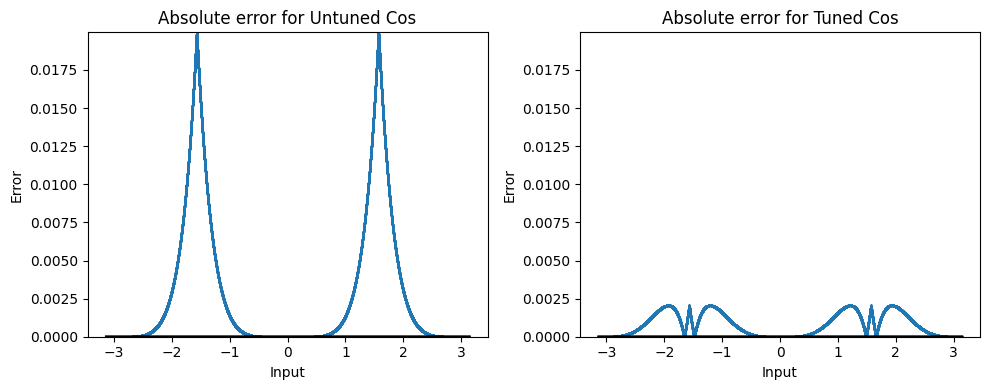}

  \caption{Absolute error for two implementation of \F{cos}:
    on the left, the un-tuned $\F{impl}_5$,
    and on the right, its tuned counterpart.
  The only difference is the coefficient for $x^4$,
    which is $\frac{1}{24}$ on the left
    and a Sollya-synthesized $0.0387248842917668$ on the right.
  }
  \label{fig:tuning}
  \Description{Error for the un-tuned cosine implementation $\F{impl}_5$
    and its  tuned counterpart.
  The un-tuned implementation has two spikes of error
  at approximately $\pm\pi/2$, with error spiking to almost 0.02 absolute error.
  On the right the error is much lower, below 0.0025.}
\end{figure}

So far, our implementation just describes a high-level strategy,
  with many important details left unspecified.
That's on purpose:
  \name allows users to express a high-level strategy
  without having to make low-level decisions.
Those low-level decisions can then be tuned by
  compiling, measuring, and tweaking the implementation
  to achieve the desired speed and accuracy.
To start this tuning cycle, a user
  first verifies that their implementation is semantically well-formed
  by calling \F{verify}.
They can then call \F{compile}
  to compile the implementation to C source code,
  and then \F{measure} to run the compiled code
  on thousands of inputs to measure both accuracy and speed,
  presenting the result in a plot.
\Cref{fig:tuning} shows the plot for $\F{impl}_5$ above.

One thing that stands out in the plot
  is the spike in error near $\frac\pi2$.
This is a consequence of using Taylor series coefficients,
  which are only accurate near $0$.
A more accurate polynomial is needed.
One of the best existing tools for that is Sollya~\cite{sollya},
  which implements state-of-the-art polynomial fitting algorithms
  such as \F{remez} and \F{fpminimax} algorithms.
Normally, a math library implementor
  would invoke Sollya to derive polynomial coefficients,
  and then copy those coefficients into their implementation.
\name, however, can automate that process using \textit{holes},
  which in \name are written $?\Impl{f(x)}{I}$,
  with the $f(x)$ and $I$ indicating the type of term to be put in the hole.
Here, we can replace the \F{approx} term
  with a more-accurate, synthesized approximation:
\[
\F{impl}_2 =
{?\Impl{\cos(x)}{\left[0, \frac\pi2\right]}}.%
\F{synthesize}(\str{remez}, \F{powers}{=}[4], \F{fixed}{=}\{0:a_0, 2:a_2\})
\]
The first argument to \F{synthesize} names the synthesis tool to use,
  and the other parameters control its behavior.
Here, we ask \name to synthesize only an $x^4$ term,
  while keeping the $x^0$ and $x^2$ terms fixed at their current values.
This takes $2.1$ seconds,
  and the new implementation, with the filled hole,
  is returned and can then be compiled and remeasured,
  with the results shown in \Cref{fig:tuning}.
\name supports all of Sollya's other polynomial fitting tools,
  and could be easily extended to call out to
  other software packages like Mathematica, Maple, or Matlab.

\name also allows the precise computation
  to be tuned using tuning parameters.
For example, when evaluating a polynomial,
  it is sometimes useful to evaluate the first few terms
  in a higher precision than later terms.
\name's optional \F{split} and \F{split\_precision} options
  allow tuning this.
For example,
  to evaluate the first two terms of the polynomial
  in double-precision arithmetic,
  but the last term in single-precision arithmetic,
  one could write:
\[
\F{impl}_1 = \F{polynomial}\left(\left[0: 1, 2: -\frac12, 4: \frac1{24}\right], \F{precision}{=}\F{fp32}, \F{split}{=}2, \F{split\_precision}{=}\F{fp64}\right)
\]
Users can likewise tune the polynomial evaluation scheme
  (using Estrin or Horner form, which matters for longer polynomials),
  how reductions are executed,
  and numerous other low-level decisions.
While tuning parameters can affect
  the speed and accuracy of the resulting code,
  they can't affect syntactic or semantics wellformedness
  and don't affect the high-level algorithm.

Moreover, \name provides a couple of special \name constructors
  that are useful for more complex tuning operations.
For example, near 0, $\cos(x)$ is exactly equal to 1.
If lots of such inputs are expected,
  it might be worth special-casing these inputs,
  like so:
\begin{align*}
& I_0 = [-0.0003, 0.0003], I = [-\pi, \pi] \\
& \F{impl}_s = \F{approx}(\cos(x), I_0, 2^{-24}, \F{polynomial}([0: 1])) \\
& \F{impl}_6 = \F{split}([I_0 \mapsto \F{impl}_s, I \mapsto \F{impl}_5]),
\end{align*}
Here $\F{impl}_s$ is the degenerate ``polynomial'' $1$,
  and $\F{impl}_6$ uses that approximation only on special case domain $I_0$.
Note that the \F{approx} term's error bound, $2^{-24}$,
  guarantees that this approximation is only used
  where it introduces no error.
Developing these split implementations can be challenging
  because of the complex interactions between speed and accuracy.
But \F{split} tests semantic wellformedness
  just like other \name operators,
  the user can experiment fearlessly,
  rely on \name to catch errors as they perform this experimentation.
Of course, \F{split} may or may not ultimately improve speed;
  only the user can ultimately know if this trade-off,
  or other tradeoffs like a slightly wider $I_0$,
  is worth it.

\subsection{Synthesizing \name Implementations}



The \name workflow described so far---%
  where an expert first crafts the high-level implementation,
  and then adjusts tuning parameters---%
  is perfect for experts with pre-existing numerics knowledge.
Less knowledgable users, however,
  benefit from the strong automated tooling
  that \name's strict type system enables.
For example, a user might want an implementation of $1 - \cos(x)$
  but not be comfortable writing it themselves.
\name allows them to get started anyway:
\[
?\Impl{1 - \cos(x)}{[-\pi, \pi]}.\F{synthesize}(\str{tds}, \str{remez})
\]
Here, the \F{synthesize} method is passed not only \F{remez},
  for finding polynomial approximations,
  but also \F{tds}, \name's \emph{type-directed synthesis} algorithm.
Type-directed synthesis works by ``reversing \name's type rules'',
  and can synthesize range reduction terms
  like the \F{left} and \F{right} terms above,
  as well as more complex \F{periodic} and \F{logarithmic} reductions.
Moreover, type-directed synthesis
  can generate multiple variations of a desired function from scratch,
  or help complete partial implementations.
The synthesized terms can then be used directly or tuned by the user.

\section{DSL and Type System}
\label{sec:dsl}

The \name DSL expresses math functions implementations
  safely, modularly, and tunably.

\begin{figure}
    \begin{grammar}
        <impl> ::=
        $\F{polynomial}([n : a, \dotsc])$
        | $\F{approx}(f(x), I, \varepsilon, <impl>)$
        \alt $\F{left}(s(x), <impl>, t(y))$
        | $\F{right}(s(x), <impl>, t(y))$
        \alt $\F{periodic}(p, <impl>, t(y, k))$
        | $\F{logarithmic}(p, <impl>, t(y, k))$
        \alt $<impl> \odot <impl>$
        | $(y = <impl>) ; <impl>$
        \alt $\F{split}([I \mapsto <impl>, \dotsc])$
        | $\F{rewrite}(<impl>, f(\dotsc) \mapsto g(\dotsc))$

    \end{grammar}
    \begin{flalign*}
    & k, n : \mathbb{Z};
      \quad a, b, c, p, \varepsilon : \mathbb{R};
      \quad I : \mathbb{R}\times\mathbb{R};
      \quad \odot \in \{ +, -, \cdot, / \}
      \quad f, g, s, t \in \text{mathematical expressions}
      &
    \end{flalign*}

    \caption{The \name grammar.
      Intervals, constants, and real-valued expressions
        are all represented symbolically.
      \F{polynomial} and \F{approx}
        are used to build polynomial and rational approximations.
      \F{left} and \F{right} remap one half of the domain
        to the other half.
      \F{periodic} and \F{logarithmic}
        do additive and multiplicative range reduction.
      \F{split} and \F{rewrite}
        to add special cases
        or tweak \name's generated code.
    }
    \label{fig:dsl}
    \Description[EBNF grammar]{
      The grammar of MegaLibm contains three main syntactic classes.
      Expressions include variables, real constants, the basic
      arithmetic operations of addition, subtraction, multiplication,
      and division, and also an array of standard mathematical functions.
      Approximations include polynomial, rational, and more complex
      multipolynomial approximations.
      Implementations include arithmetic operations; compositions;
      polynomials; inflection operators that remap part of the input domain;
      periodic and logarithmic argument reduction;
      and the generic split, rewrite operations.
    }
\end{figure}

\begin{figure}
\begin{mathpar}
  \small
\inferrule{
  p(x) = \sum_i a_i x^{n_i} \\
  I = [-\infty, \infty]
}{
  \F{polynomial}([n_i : a_i]) : \Impl{p(x)}{I}
}
\and
\inferrule{
  e : \Impl{g(x)}{J} \\
  I \subseteq J \\\\
  \forall x \in I,\;|f(x)-g(x)| < \varepsilon
}{
  \F{approx}(f(x), I, \varepsilon, e) : \Impl{f(x)}{I}
}
\and
\inferrule{
  e : \Impl{f(x)}{[m, b]} \\
  m = (a + b) / 2 \\\\
  \forall x \in [a, m],\;s(x) \in [m, b]
  \land t(f(s(x))) = f(x)
}{
  \F{left}(s(x), e, t(y)) : \Impl{f(x)}{[a, b]}
}
\and
\inferrule{
  e : \Impl{f(x)}{[a, m]} \\
  m = (a + b) / 2 \\\\
  \forall x \in [m, b],\;s(x) \in [a, m]
  \land t(f(s(x))) = f(x)
}{
  \F{right}(s(x), e, t(y)) : \Impl{f(x)}{[a, b]}
}
\and
\inferrule{
  P = [0, p] \\ e : \Impl{f(x)}{P} \\\\
  \forall x \in P,\;t(f(x), k) = f(x + p k)
}{
  \F{periodic}(p, e, t(y, k)) : \Impl{f(x)}{[-\infty, \infty]}
}
\and
\inferrule{
  P = [p^{-1/2}, p^{1/2}] \\
  e : \Impl{f(x)}{P} \\\\
  \forall x \in P,\;t(f(x), k) = f(p^k x)
}{
  \F{logarithmic}(p, e, t(y, k)) : \Impl{f(x)}{[0, \infty]}
}
\and
\inferrule{
  p : \Impl{f(x)}{I} \\
  q : \Impl{g(x)}{I}
}{
  p \odot q : \Impl{f(x) \odot g(x)}{I}
}
\and
\inferrule{
  p : \Impl{f(x)}{I} \\
  q : \Impl{g(y)}{J} \\\\
  \forall x \in I, \; f(x) \in J \land g(f(x)) = h(x)
}{
  (y = p) ; q : \Impl{h(x)}{I}
}
\and
\inferrule{
  e_i : \Impl{f(x)}{I_i} \\
  \bigcup_i I_i = I
}{
  \F{split}([ I_i \mapsto e_i ]_i) : \Impl{f(x)}{I}
}
\and
\inferrule{
  e : \Impl{f(x)}{I} \\\\
  \forall \vec{x} \in \mathbb{R}^n,\; a(\vec{x}) = b(\vec{x})
}{
  \F{rewrite}(e, a \mapsto b) : \Impl{f(x)}{I}
}
\end{mathpar}
\caption{\name's type rules.
  Each type rule's antecedents
   include first a set of syntactic wellformedness conditions,
   which guarantee that an implementation implements
   the correct function on the correct range,
   and then, on the next line, a set of semantic wellformedness conditions,
   which guarantee the correctness of the implementation
   over the reals.
}
\label{fig:typerules}
\Description{\todo{desc}}
\end{figure}

\subsection{Syntax and Semantics}

The most important terms in the \name DSL have type \Impl{f(x)}{I},
  where $f$ is the \textit{target function} in argument $x$,
  and the \textit{domain} $I$ is an interval $[x_{lo}, x_{hi}]$
  of real values that $x$ is expected to take on.
Both the function and the interval
  are represented symbolically by an AST
  containing operators (addition, subtraction, multiplication, and division),
  functions (square root, trigonometric, and exponential functions),
  constants (explicit and symbolic),
  and the variable $x$.%
\footnote{
This symbolic representation of intervals is important
  for implementing functions like logarithm
  where domains like $[2^{-1/2}, 2^{1/2}]$
  need to be represented exactly.
}
Typically, the user does not need to explicitly write types,
  which for each term can be computed from its arguments.

The full grammar of \name is given in in \Cref{fig:dsl}.
The \F{polynomial} term is the only leaf term,
  while all other terms combine other implementation terms.
To guarantee safety, the \name DSL is typed,
  and each type rule is split
  into \emph{syntactic} and \emph{semantic} wellformedness.
Syntactic wellformedness merely propagates
  target functions and domains.
Semantic wellformedness checks
  that the implementation would work correctly over the real numbers
  and guarantees the absence of mathematical mistakes, sign flips, and typos.
Formally, every MegaLibm type describes a real-valued function, while
  each term denotes a set of real-valued functions; the set structure
  is required to support \F{approx}.
Semantic well-formedness ensures that the function represented by the type is
  included in the set represented by the term.
The full type system for \name terms
  is given in \Cref{fig:typerules};
  checking semantic wellformedness is discussed in \Cref{sec:check}.

\subsection{Approximation Terms}

Dozens of function approximation schemes exist
  including Taylor series, Chebyshev approximations,
  Remez exchange, including its variations for non-Haar spaces,
  LLL for rounding polynomial coefficients,
  Caratheodory-Fejer approximations, Pad\'e approximations,
  and many others, including active research
  such as the RLibM project~\cite{rlibm1,rlibm2,rlibm3}.
\name must therefore provide the flexibility
  to use any polynomial or other function approximation
  while still guaranteeing safety.
\name accomplishes this through its combination
  of \F{approx} and \F{polynomial} terms.%
\footnote{\name also provides shorthands
  for some common non-polynomial approximations,
  which can be thought of as combinations of \F{polynomial} terms.}

A \F{polynomial} term takes a list of powers and coefficients,
  and is typed as an implementation of that polynomial
  over the full range $[-\infty, \infty]$.
Rational polynomials and other more-complex polynomial forms
  can be defined using constructions such as:%
\footnote{This particular term is Bhaskara I's famous
  approximation of the sine function.}
\[
\F{polynomial}([1: 16\pi, 2: -16])
/
\F{polynomial}([0: 5\pi^2, 1: 4\pi, 2: -4]).
\]
The polynomial coefficients are provided directly by the user,
  providing total flexibility
  to compute those coefficients using whatever method seems best.
Short \F{polynomial} terms can be used
  to represent constants or simple linear approximations
  for special cases.
The \F{polynomial} coefficients are symbolic
  and are rounded to the appropriate precision during compilation.

There are many ways to evaluate any given polynomial
  in floating-point arithmetic.
The precision is configurable using tuning parameters,
  as is the polynomial evaluation form (Horner or Estrin).
The \F{split} tuning parameter also enables
  a limited but common form of mixed-precision evaluation.
And \F{rewrite} terms (described below)
  can be used to leverage polynomial factoring.
These choices are part of tuning and don't affect type checking.

Function approximations
  must then be wrapped with an \F{approx} term
  to cast them to the appropriate \F{Impl} type.
The \F{approx} term
  declares the target function being approximated,
  the domain over which it is being approximated,
  and the maximum absolute approximation error over that domain
  (see \Cref{fig:typerules}).
This serves to catch typos like mixing up the signs of coefficients
  or entering the wrong powers of $x$.
When synthesizing function approximations (see \Cref{sec:synthesis}),
  the error bound is automatically filled in by the synthesis tool.
The \F{approx} term does not generate any code
  and is present solely for safety and type checking.

\subsection{Range Reduction Terms}

Range reduction and reconstruction algorithms
  remap inputs and outputs
  from one domain to another, typically smaller, domain.
These algorithms are specific to the function being implemented,
  and state-of-the-art math libraries use
  a great variety of different approaches.
Moreover, range reduction and reconstruction
  typically contain a mix of multiple steps,
  iteratively increasing the domain the function is evaluated on.
\name must therefore provide a flexible and modular framework
  for range reduction and reconstruction.

\name relates range reduction and reconstruction
  to identities of the target function.
For example, a \F{sin} implementation
  might split the sign and magnitude of $x$,
  computing \F{sin} of the magnitude
  and then adding back the sign afterwards.
In \name, this range reduction and reconstruction algorithm
  is seen to derive from the identity $\sin(x) = -\sin(-x)$.
This identity implies
  that \F{sin} can be computed
  either by evaluating it directly
  or by \emph{reducing} the input $x$ to $-x$,
  then evaluating \F{sin},
  and then \emph{reconstructing} the output $y$ to $-y$.
An implementation can make use of this
  by using the $-\sin(-x)$ implementation for negative $x$
  and the $\sin(x)$ implementation for positive $x$,
  and thereby guaranteeing that core \F{sin} approximation
  is only called on positive inputs.
More abstractly, each range reduction and reconstruction algorithm
  derives from a pair $s$/$t$ of functions
  such that the identity $t(f(s(x))) = f(x)$ holds.
Importantly, such $s$/$t$ pairs can be composed:
  if $t_1(f(s_1(x))) = f(x)$ and $t_2(f(s_2(x))) = f(x)$,
  then by substitution the composition
  $t_1(t_2(f(s_2(s_1(x))))) = f(x)$
  holds as well.
Range reductions therefore form a monoid,
  which allows composing simple range reductions
  to construct more complex ones.

This abstract model is expressed in \name
  via four range reduction operators:
  \F{left}, \F{right}, \F{periodic}, and \F{logarithmic}.
The \F{left} and \F{right} \name operators
  simply correspond to a branch
  that applies $t(f(s(x)))$ on one half
  of the domain and $f(x)$ on the other half.%
\footnote{
\F{left} and \F{right} could be made more general:
  the midpoint $m$ could be user-provided.
This extension would not substantially change the design of \name,
  but we didn't find a need it for any of the functions
  that we wanted to implement.
}
For example, \F{right} has the type signature:
\[
\inferrule{
  e : \Impl{f(x)}{[a, m]} \\
  m = (a + b) / 2 \\\\
  \forall x \in [m, b],\;s(x) \in [a, m]
  \land t(f(s(x))) = f(x)
}{
\F{right}(s(x), e, t(y)) : \Impl{f(x)}{[a, b]}
}
\]
The syntactic condition requires that $e$ can handle
  inputs from the left half of the domain,
  while the semantic condition
  requires that $s$ reduces inputs
  from the right half of the domain to its left half
  and that the $s$/$t$ identity holds.
This prevents mathematical errors in either $s$ or $t$
  and ensures that the valid domains for each computation
  are propagated correctly.
Like the rest of the \name type rules,
  these semantic conditions
  are purely mathematical and can be checked
  statically.

\name also provides \F{periodic} and \F{logarithmic} operators,
  which correspond to repeatedly composing a range reduction with itself.
For example, the \F{exp} function satisfies the identity
  $\exp(x) = \exp(x + a) / e^a$,
  yielding $s(x) = x + a$ and $t(y) = y / e^a$.
By repeatedly applying $s(x)$, or its inverse,
  any input can be reduced to the range $[0, a]$.%
\footnote{Of course,
  implementations don't actually contain a \F{while} loop
  that repeatedly adds or subtracts.
Implementations instead use algorithms like Cody-Waite reduction
  to compute a high-precision modulus operation.}
More generally, this corresponds to the $k$-times-iterated identity
  $t^k(f(s^k(x))) = f(x)$.
In \name, two such $k$-times-iterated identities are supported:
  $s(x) = x + a$ via \F{periodic}
  and $s(x) = x \cdot a$ via \F{logarithmic}.
In each case, the user supplies the function $t^k(x)$, written $t(x, k)$,
  instead of just $t(x)$.
Allowing the user to specify $t^k(x)$ makes \name quite general;
  for example, the same operator, \F{periodic},
  is used to define not just the periodic function \F{sin}
  but also the exponential function \F{exp}:
\[
  \F{exp\_impl} = \F{periodic}(\log(2), \dotsc, \F{ldexp}(y, k))
\]
Nesting range reduction operators composes their $s$/$t$ pairs.
The individual range reductions can therefore be checked independently,
  but safety is guaranteed for the range reduction as a whole.
This makes the implementation easier to read
  and allows constructing and testing range reductions incrementally.
While not strictly speaking a range reduction algorithm, the \name
  composition operation $(y = p);q$ is also commonly used to change input
  domains by computing intermediate values.
This expression represents the function $q \circ p$, reversing the order of
  composition to match traditional programming notation, and also introduces
  the variable y, which can be used within q.
It can be, and often is, freely mixed with other range reduction steps.

Just like \F{polynomial} terms, range reductions can be tuned.
The \F{periodic} and \F{logarithmic} operators do not specify
  working precision or the specific reduction algorithm.
For example, to implement $\cos(x)$ over a large range,
  one might want to use Payne-Hanek reduction,%
\footnote{The current \name prototype
  does not implement Payne-Hanek reduction,
  though we hope to add it in the future.}
  while $\cos(x)$ over a smaller range might use Cody-Waite reduction
  and $\cos(\pi x)$ might use simple division;
  the \F{method} tuning parameter can specify this
  without changing the real-number semantics of the reduction.
Likewise, the precision of reduction and reconstruction in \F{left} and \F{right}
  can be specified by tuning parameters.
The floating-point reduction bounds, periods, and multi-precision constants
  are all computed automatically by \name during compilation
  and do not have to be specified by the user.

\subsection{Tuning for Speed and Accuracy}

\name functions have fixed semantics over the real numbers,
  but their behavior when compiled is more complex:
  many low-level decisions become relevant
  for the speed and accuracy of the resulting code.
In \name, these low-level decisions are accessible
  as optional tuning parameters,
  which don't affect the syntax or semantics of \name terms
  but do affect their compilation to C and thereby speed and accuracy.
This design guarantees safety
  by ensuring that tuning parameters can be freely adjusted by users
  and also separates high-level and low-level reasoning,
  creating a form of modularity.
Moreover, tuning parameters can be freely added as \name is extended,
  giving users intimate control over speed and accuracy,
  a property we term \emph{tunability}.

A complete list of \name's current tuning parameters
  is given in \Cref{fig:parameters}.
All \name terms%
\footnote{Except no-op terms like \F{approx}.}
  have a \F{prec} tuning parameter
  to select the working precision of intermediate operations.
For example, in \F{left} terms, this working precision
  is used for the evaluating
  the reduction and reconstruction functions $s(x)$ and $t(y)$.
\name supports the standard \F{fp32} and \F{fp64} precisions
  as well as compensated ``double-double'' operations,
  which are used in many math libraries.
Double-double operations compile to calls
  to a custom library based on the CRLibm project~\cite{crlibm}
  and David Bailey's Fortran library~\cite{dbailey},
  which provides the standard addition and subtraction operations,
  multiplication and division operations based on \F{fma},
  and a square root operation extracted from Sun's fdlibm library~\cite{fdlibm}.

Other tuning parameters are specific to individual terms.
For example, a \F{polynomial} term
  evaluates a polynomial in Horner form by default,
\[
a_0 + x\cdot(a_1 + x\cdot(a_2 + x\cdot(a_3 + x\cdot(a_4 + x\cdot(a_5 + x\cdot(a_6 + x\cdot a_7)))))))
\]
However, other evaluation schemes exist.
Estrin form uses a tree structure for evaluation:
\[
((a_0 + x \cdot a_1) + x^2(a_2 + x\cdot a_3)) + x^4((a_4 + x\cdot a_5)
+ x^2(a_6 + x \cdot a_7))
\]
This is typically mildly less accurate than Horner form,
  but can be faster on super-scalar processors because
  it has a critical path of length $O(\log N)$, not $O(N)$,
  where $N$ is the number of polynomial terms.%
\footnote{
Moreover, on processors with vector units,
  some of the operations can be efficiently vectorized.
}
It is can also be valuable to split out leading terms, like so with three split terms:
\[
a_0 + (x \cdot a_1 + (x^2 \cdot a_2 + x^3(a_3 + x(a_4 + x(a_5 + x(a_6 + x\cdot a_7)))))))))
\]
All of these options are available to the user
  via the \F{method} and \F{split} tuning parameters on \F{polynomial},
  and when leading terms are split out,
  they can be computed in a different (typically higher) precision
  using the \F{split\_prec} tuning parameter.
\F{periodic} likewise has various tuning parameters
  for adjusting how the additive range reduction is actually done.
If Cody-Waite range reduction is chosen,
  the \F{cw\_bits} and \F{cw\_len} parameters define
  the precision of the various constants used,
  which affects speed and accuracy in non-trivial ways.
Some tuning parameters do additional checks
  and raise errors if those checks fail;
  for example, if a \F{polynomial} term is asked to \F{split}
  more terms than the polynomial actually has,
  or if the \F{split} argument is negative.
Because these mistakes are easy to diagnose and fix,
  we think of these as syntactic, not semantic, errors.
Tuning parameters also provide a measure
  of extensibility and future-proofing to \name.
New algorithms, or variations of existing algorithms,
  are easy to add to \name as tuning parameters.
As long as the tuning parameters have sensible defaults,
  existing implementations would not be affected.

\begin{figure}
  \begin{tabular}{l|l|l}
    \bf Operation & \bf Parameter & \bf Effect \\ \hline
    All       & \F{prec}      & Precision for operations: \F{fp32},
    \F{fp64}, \ldots \\
    \F{polynomial}    & \F{method}    & Polynomial evaluation method:
    \F{horner}, \F{estrin}, \ldots \\
    & \F{split}     & Number of leading terms to evaluate separately \\
    & \F{split\_prec}     & Precision to evaluate leading terms in \\
    \F{periodic}  & \F{method}    & Reduction method: \F{naive}, \F{cody-waite}, \ldots \\
     & \F{cw\_bits}  & Bits per element in the Cody-Waite constant \\
     & cw\_len   & Elements in the Cody-Waite constant
  \end{tabular}
  \caption{Tuning parameters for MegaLibm operations.
    Each impacts the generated code coming,
      but does not affect syntactic or semantic wellformedness.
    Additional tuning parameters can therefore be
      easily added without affecting the \name core
      or the behavior of existing \name implementations.
    Some parameter settings are invalid and raise error:
      \F{split}ting more terms than a polynomial contains,
      or using \F{cody-waite} reduction without specifying
      a \F{cw\_bits} or \F{cw\_len}.}
  \label{fig:parameters}
  \Description[Parameter table] {
    Table containing optional parameters to MegaLibm operations.
    All operations accept prec, which sets the implementation precision.
    Scheme accepts method, which sets the polynomial evaluation method.
    Scheme accepts split, which sets the number of leading terms to evaluate
      separately.
    Periodic accepts method, which sets the reduction method.
    Periodic accepts cw\_bits, which sets the number of bits per element when
      using Cody-Waite.
    Periodic accepts cw\_len, which sets the number of elements when using
      Cody-Waite.
  }
\end{figure}

Some accuracy and performance tweaks, however,
  cannot be easily represented with tuning parameters
  because the changes they make to the implementation are too invasive.
\name provides the \F{split} and \F{rewrite} operators to handle these cases.
The \F{split} operator
  allows an implementation to use different approximations
  for different portions of the input domain,
  such as small-angle approximations in \F{sin}.
An implementation can also use \F{split}
  to switch reduction algorithms or polynomial schemes
  depending on the input,
  such as using naive range reduction on inputs near 0
  and Cody-Waite range reduction on inputs further away
  when implementing \F{sin}.

The \F{rewrite} operator
  allows rewriting one mathematical expression into another,
  presumably with higher accuracy.
For example, in Sun \texttt{fdlibm}'s \F{log} implementation,
  the remapping $s=\frac{f}{2+f}$ is performed
  so that $2s = f - s \cdot f$.
It turns out that $s$ has rounding error but $f$ is exact,
  so replacing $2s$ by $\F{fma}(-s, f, f)$ improves accuracy.
A \F{rewrite} operator allows the user to make this replacement in one place.
To help support complex rewrites,
  the \name composition operator $(y = p) ; q$
  introduces a name, $y$, for the intermediate value,
  and \F{rewrite} can refer to this name.

Both \F{split} and \F{rewrite} have semantic wellformedness rules
  that guarantee that their use preserves
  the mathematical correctness of the implementation,
  so even when making such invasive changes to an implementation,
  a \name user can still rely on the \name type system
  to rule out typos and mathematical mistakes.

\section{The \name Toolchain}
\label{sec:tools}

\name is typically used via a Jupyter notebook.
The user constructs the AST directly,
  with each term in the DSL having a corresponding AST constructor
  and returning an object of type \F{Impl}.
Once a \name user writes a complete implementation,
  that implementation needs to be type-checked
  and can then be compiled to C
  to measure its speed and accuracy.

\subsection{Type Checking}
\label{sec:check}

Terms in \name must be checked
  for both syntactic and semantic wellformedness.
While syntactic wellformedness can typically be checked
  using a simple top-down recursive type checking procedure,
  semantic equivalence requires reasoning about real numbers
  and therefore requires more complex techniques.
\name therefore performs syntactic type-checking automatically,
\footnote{Since \name intervals are represented symbolically,
  syntactic wellformedness can require reasoning about real numbers.
The only tests necessary, however, are comparisons.
Our prototype therefore just uses a high-precision evaluation,
  but deferring these checks to the semantic type-checking phase
  would also be possible.}
  as terms are constructed,
  but defers semantic wellformedness checks
  to a separate \F{check} method.
This \F{check} method traverses the \name term,
  gathers all semantic wellformedness constraints
  given by the type rules in \Cref{fig:typerules},
  and attempts to prove each.


Each constraint takes the form of a comparison or equality
  between two real-valued expressions
  with variables drawn from intervals.
Unfortunately, determining equality for arbitrary real-valued expressions
  is known to be hard---
  dependent on unproven mathematical conjectures~\cite{api-reals},
  and possibly undecidable.
\name therefore cannot provide a complete decision procedure for this problem;
  instead, it provides access to several different backends,
  which differ in their soundness and completeness:
  \F{egg}, \F{sympy}, \F{dirtyinfnorm}, and \F{sampling}.
New verification tools such as Mathematica or Maple
  could be easily added in the future.

The \F{egg} backend can prove equalities (but not inequalities)
  using the egg e-graph library~\cite{egg}
  and a custom set of rewrite rules
  initially drawn from Herbie~\cite{herbie}
  but with unsound rules removed.
To prove two expressions equal,
  both are added to a single e-graph,
  and then the rewrite rules are run
  for a user-configurable number of iterations
  or until the e-graph reaches a user-configurable size bound.
If the e-nodes corresponding to the two expressions
  are in the same e-class at the end of the iteration,
  the expressions are equal; otherwise, the \F{egg} backend reports an error.
The \F{egg} backend is fast and, to the best of our knowledge, sound,
  but is unable to prove some more complex identities.
The \F{sympy} backend uses the \F{simplify} method
  from Python's SymPy library.
Sympy internally uses a normalization procedure instead of rewrites,
  and is mature and well-tested;
  it can prove both equalities as well as some inequalities.
Like the \F{egg} backend, the \F{sympy} backend is sound
  (at least, assuming no bugs in SymPy)
  but not complete.

\name also provides some unsound checking algorithms.
The \F{dirtyinfnorm} backend checks comparisons,
  such as the \F{approx} operator's wellformedness condition.
It uses interval arithmetic at evenly-spaced points along the domain
  as well as some form of first-order search
  to find possible minima and maxima.
It typically yields tight bounds, but in theory is not sound.
The \F{sampling} backend can verify any kind of condition
  by sampling inputs in the domain,
  evaluating all terms in high precision,
  and testing all conditions up to some tolerance;
  the precision and tolerance can be configured by the user.
Since the user chooses the set of verification tools to use,
  the soundness and completeness of the result
  is ultimately up to the user.

\subsection{Compilation}

Each term in the \name DSL can be compiled to C
  by invoking its \F{generate\_c} method.
Internally, this occurs in two steps:
  first, converting the \name term into an intermediate representation,
  which requires lowering all real-number constants to floating-point ones,
  and then converting that intermediate representation into C.
The intermediate representation
  takes the form of an ordered sequence of untyped computation nodes
  linked together in a graph,
  analogous to a ``sea of nodes'' SSA IR.
Each node has a fixed count of input and output variables,
  plus named configuration parameters that typically reflect
  tuning parameters in the \name DSL.
The first input and output parameter is special:
  it always represents the transformed input $x$ or output $y$.
For instance, the \texttt{cast} block
  takes a single input, produces a single output,
  and has a \F{c\_type} parameter to define the underlying type.
The intermediate representation
  has no representation of control flow:
  all IR nodes are computed in order, one after another.
This is possible
  because branches in math function implementations
  are typically short,
  more like conditional moves than true branches.
Avoiding control flow
  dramatically simplifies the intermediate representation.

\begin{figure}
  \begin{tabular}{l|l|l|l}
    Node       & Input(s) & Output(s) & Additional Arguments \\
    \hline
    \F{cast}        & 1        & 1         & c\_type \\
    \F{cody\_waite} & 1        & 2         & period, bits\_per, entries \\
    \F{decompose}   & 1        & 2         & \\
    \F{expression}  & any      & 1         & expr \\
    \F{estrin}      & 1        & 1         & monomials, coefficients, split \\
    \F{horner}      & 1        & 1         & monomials, coefficients, split \\
    \F{if\_less}    & 1        & 1         & bound, t\_val, f\_val \\
    \F{mod\_switch} & 2        & 1         & mod\_to\_blocks \\
    \F{set\_exp}    & 2        & 1         & \\
    \F{additive}    & 1        & 2         & period \\
    \F{split\_dom}  & 1        & 1         & dom\_to\_block
  \end{tabular}

  \caption{The nodes used in \name's intermediate representation.
    Inputs and outputs to nodes always result in C-level variable names.
    Additional arguments are mandatory and vary in type.
 }
  \label{fig:blocks}
  \Description[Table of block functions]{
    TODO
}
\end{figure}

\Cref{fig:blocks} lists all of \name's intermediate representation nodes.
The \F{cast} block handles casts between \name's supported number representations.
The \F{decompose} and \F{set\_exp} nodes convert between a floating-point number
  and the integer fields that comprise it;
  they are analogous to the C standard \F{frexp} and \F{ldexp} functions,
  but directly use bit tricks because \name assumes
  that \F{NaN} and infinite values are handled separately.
The \F{estrin} and \F{horner} nodes
  implement the corresponding \F{polynomial} evaluation schemes.
The \F{additive} and \F{cody\_waite} nodes
  handle \F{periodic} terms with \F{naive} or \F{cody-waite} methods;
  the \F{cody\_waite} term has additional arguments
  to configure the high precision constant.
The \F{expression} node is used for reduction and reconstruction functions
  and evaluates arbitrary mathematical expressions by the corresponding C operator.
Finally, \name's intermediate representation
  has several specialized conditionals.
The \F{if\_less} node handles conditionals in \F{left} and \F{right} terms;
  specializing \F{if} to comparisons avoids
  having to explicitly represent boolean operations.
However, the \F{mod\_switch} node
  is a specialized conditional for branching
  on the lower bits of an integer constant $k$
  using C's \F{switch} operator,
  which is useful in \F{periodic} reconstruction functions.
\F{split\_dom} uses a chain of \F{if}/\F{else} conditions
  to handle the interval testing in \F{split} terms.
These specialized conditional nodes
  are automatically inserted by \name,
  and are important for achieving good performance.

To compile an implementation,
  its terms are traversed in post-order
  and each term generates nodes following a template.
Terms do not examine the compiled form of other nodes,
  except in the case of \F{rewrite}.
Input and output types (\F{fp64}, \F{fp32}, or \F{dd} for double-double)
  are tracked during compilation
  and casts are automatically inserted
  to account for \F{prec} tuning parameters.
After all the nodes are generated,
  names are chosen for each intermediate variable
  and C code is produced for the sequence of nodes.
While \name currently only supports C,
  in the future we hope to develop additional backends.
The IR-based architecture should simplify that extension.
\name does not perform any kind of optimization to the IR,
  and frequently generates code with common subexpressions
  and dead code.
\name therefore relies on the C compiler's optimizer
  to eliminate redundancy;
  because the IR node typically contains simple floating-point operations,
  simple control flow, and \F{always\_inline} functions,
  the C compiler's optimizer is typically sufficient
  and the resulting assembly code is acceptably fast.

\subsection{Measurement}
\label{sec:measure}

Once a \name implementation is compiled to C,
  it can be run on randomly sampled points
  to estimate its speed and accuracy.
Importantly, this measurement step is integrated into the Jupyter workflow,
  facilitating the user rapidly compiling, testing, and tweaking
  their implementation.

To measure a compiled implementation,
  \name generates and compiles a series of drivers.
Each driver randomly generates inputs
  uniformly distributed over an interval;
  by default, the domain of the implementation itself is used,
  but the user can optionally provide a subset of the domain,
  which is useful for ``zooming in'' on inputs of interest,
  such as when developing \F{split} terms.
After sampling points in the domain and saving them in memory,
  \name then invokes the implementation on each point in turn,
  saving the results for later analysis.
The total execution time is saved as a measure of the implementation's speed.
Naturally, execution speed can vary
  depending on the execution context, machine state,
  and the specific distribution of inputs used.
However, since math function implementations are typically
  arithmetic-heavy and control-flow-light,
  simple measurement over uniformly sampled inputs
  tends to provide a meaningful measure of speed.

To measure the implementation's accuracy,
  each output is compared to an oracular result computed by MPFR.
The number of points is configurable by the user.
For each output, we compute the absolute error
  compared to the oracular result
  and plot all of the outputs in a dot plot,
  such as the one in \Cref{fig:tuning}.
Note that this produces an error plot,
  not a worst-case analytic bounds.
While analytic bounds are the ultimate goal
  of math function implementation,
  it is often important, while a function is being developed,
  to focus on how error varies with input
  and how common inaccurate inputs are
  (which often provides a hint how to fix them).
Verification of math function accuracy
  is still an active area of research~\cite{verified-math-h,satire}.
As the tools in this area mature,
  we plan to integrate them into \name.

Debugging speed or accuracy is helped by the fact
  that \name implementations are modular.
This means that users can typically measure the accuracy
  of individual pieces---for example,
  just the function approximation,
  or just one of several range reductions---%
  to get insight into what specific step is inaccurate or slow.
\name's tuning parameters then provide a set of options
  to achieve greater speed or accuracy for that step.
For example,
  a user might find that their implementation is inaccurate on a certain input.
They would then proceed to test, say, their core polynomial approximation,
  and if that is accurate,
  they may then try increasing the precision
  of a range reduction step.
Key to this incremental workflow
  is the fact that individual implementation steps
  are easy to separate in the \name DSL
  and that the type system ensures safety,
  avoiding misleading mistakes
  during the testing and tuning loop.

\section{Synthesis in the \name DSL}
\label{sec:synthesis}

\name's safety, modularity, and tunability
  also makes it a good target for synthesis.
To that end, \name provides easy access
  to several synthesis algorithms
  to automatically generate part or all
  of a \name implementation for a given mathematical function.

\subsection{Synthesizing Approximations}

Math library implementors already commonly use
  specialized tools to compute polynomial approximations.
Popular tools include Maple, Mathematica, Matlab, and Sollya,
  and the topic is an area of ongoing research~\cite{rlibm1}.
Sollya in particular is free, open source, and implements
  multiple algorithms including Taylor series,
  Chebyshev approximation, Remez exchange,
  and shortest-vector search.
It also has a rigorous approach to safety,
  making it a good match for \name.
\name therefore provides access to Sollya
  for synthesizing polynomial approximations.

To synthesize \name implementations,
  the user uses \emph{hole} terms.
Holes are typed;
  a hole for type \Impl{f(x)}{I} is written $?\Impl{f(x)}{I}$.
Implementations then have a \F{synthesize} method,
  which takes in a list of synthesis tools
  and attempts to apply each tools to each hole in turn,
  returning a list of synthesized implementations.
This \F{synthesize} method provides flexibility for users:
  they can synthesize each hole individually,
  passing custom tuning parameters for each one,
  or construct a partial implementation
  containing potentially multiple holes
  and then calling \F{synthesize} with a generic set of options.
Incorporating synthesis into \name
  reduces errors from copying functions and coefficients between tools,
  and supports \name's interactive workflow.
Moreover, the \name synthesis method is designed for extensibility.
Both Maple and Mathematica, for example,
  can synthesize rational function approximations to a function.
Since we do not have access to either (pricey) package,
  \name does not currently have the ability to call them,
  but these algorithms would be easy to add to \name
  and then invoke via \F{synthesize}.

To invoke Sollya, users request
  the \F{taylor}, \F{chebyshev}, \F{remez}, and \F{fpminimax} synthesis tools.
These tools synthesize
  $\F{approx}(f(x), I, \varepsilon, \F{polynomial}(C))$ terms,
  where the target function $f(x)$ and interval $I$
  come from the hole's type,
  while the error bound $\varepsilon$
  and coefficient list $C$ are computed by Sollya.
The user can customize the polynomial
  by passing arguments to \F{synthesize},
  such as specifying the number of polynomial terms,
  the powers of $x$ to use,
  the precision of the coefficients (for \F{fpminimax}),
  or any number of fixed initial terms.
If the function is known to be odd or even (see below),
  Sollya synthesizers such as \F{remez} and \F{fpminimax}
  automatically attempt to synthesize coefficients
  only for the odd or even powers
  (configurable by the user).
If Sollya fails to synthesize an approximation,
  which happens in a variety of complicated and hard-to-predict cases,
  \name attempts minor fix-ups,
  including increasing the working precision
  or slightly expand the input domain.
Users can also provide lists for parameters
  such as the number of polynomial terms
  to synthesize multiple alternative implementations
  for each hole.
The same occurs if the user
  specifies multiple function approximation tools
  such as both \F{chebyshev} and \F{remez}.

\subsection{Type-directed Synthesis}

Function implementations typically require
  range reductions as well as function approximations
  in order to be competitive.
While expert users may prefer to design range reductions by hand,
  less-expert users benefit from automation.
\name thus provides a type-directed synthesis algorithm,
  invoked as \F{tds},
  to automatically suggest range reduction and reconstruction steps.

The basic idea of \F{tds} is to ``reverse the type rules'',
  using the type of the requested hole to determine
  which range reductions and reconstructions are possible.
For example, consider synthesizing an \Impl{1 - \sin(x)}{[-\pi, \pi]},
  via a \F{left} operator.
This requires synthesizing matching functions $s$ and $t$
  such that $t(1 - \sin(s(x))) = 1 - \sin(x)$,
  where moreover $s([-\pi, 0]) \in [0, \pi]$.
One such mapping is $s(x) = -x$ and $t(y) = 2 - y$,
  which then transforms the initial term into
\[
\Impl{1 - \sin(x)}{[-\pi, \pi]} \leadsto
\F{left}(-x, \Impl{1 - \sin(x)}{[0, \pi]}, 2 - y).
\]
The expanded implementation has another hole,
  which is expanded again by \F{tds},
  repeating until a full term is built.
Typically the user invokes \name's synthesis method
  with both \F{tds} and some polynomial synthesis tool like \F{remez},
  which is used to fill in leaf terms in the implementation.
The type rules, especially semantic wellformedness,
  heavily constrain the possible implementations
  and therefore cut down the search space.

The key challenge in this type-directed synthesis approach
  is efficiently finding $s$/$t$ pairs
  for arbitrary target functions.
The \F{tds} tool leverages a key observation about these pairs:
  the reduction operation $s(x)$
  is typically one of a small number of operations
  (flips, shifts, and scales)
  while the reconstruction operation $t(y)$
  can be quite complex and specific
  to the function being implemented.
\name exploits this asymmetry
  by considering a fixed set of reduction templates $s(x)$
  and attempting to synthesize a matching $t(y)$ for each.
For example, to synthesize the term
  $\F{left}(-x, \dotsc, y)$,
  \name would attempt to instantiate
  the \F{flip} template $s(x) = a - x$
  and then find a $t(y)$ such that
  $t(1 - \sin(a - x)) = 1 - \sin(x)$
The full set of templates can be found in \Cref{fig:templates}.

\begin{figure}
  \begin{tabular}{lll}
    \bf Template & $\mathbf s(x)$ & \bf Generated term \\ \hline
    $\F{flip}(a)$ & $a - x$ & \F{left} and \F{right} \\
    $\F{shift}(a)$ & $a + x$ & \F{left}, \F{right}, and \F{periodic} \\
    $\F{scale}(a)$ & $a \cdot x$ & \F{left}, \F{right}, and \F{logarithmic} \\
  \end{tabular}
  \caption{Reduction templates for \name terms,
    including the template name,
    the reduction function $s(x)$,
    and the \name terms that can use such reductions
  }
  \label{fig:templates}
  \Description[Templates table] {
    A table containing all of \name's reduction templates:
    $\F{flip}(a)$ terms correspond to horizontal reflections
    and generate \F{left} and \F{right} terms,
    $\F{shift}(a)$ terms correspond to horizontal shifts
    and generate not only \F{left} and \F{right} terms
    but also \F{periodic} terms.
    Similarly, $\F{scale}(a)$ correspond to horizontal scaling
    and generate \F{logarithmic} terms along with \F{left} and \F{right}.
  }
\end{figure}

To find the reconstruction function for each possible template,
  \name uses equivalence graphs.
Equivalence graphs explore and compactly store
  many similar, equivalent programs~\cite{egraphs},
  and \name uses the popular \F{egg} library~\cite{egg}
  (via its Python interface \F{snake\_egg})
  for working with e-graphs.
When using \F{egg},
  the user typically provides a starting expression
  and a set of rewrite rules,
  and \F{egg} then ``grows'' the e-graph,
  exploring the space of expressions
  equivalent to the starting point
  and reachable via the rewrite rules.
\name initializes the e-graph with the term $\F{thefunc}(x)$,
  and rewrite rules including
  sound mathematical rewrite rules,
  the function definition $\F{thefunc}(x) \leftrightarrow f(x)$,
  and a rewrite rule of the form $\F{thefunc}(s_T(x)) \leadsto T$
  for each reduction template $T$.
For example, the $\F{flip}(a)$ template
  generates the rewrite rule $f(a - x) \leadsto \F{flip}(a)$.
Given a starting point like $\F{thefunc}(x) = \sin(x)$,
  the e-graph will find an equivalent form like $-\sin(0 - x)$
  using the mathematical rewrite rules,
  convert it into $-\F{thefunc}(0 - x)$ via the definition rule,
  and then rewrite it into $-\F{flip}(0)$
  using the reduction template rewrite rule.
Note that this final form does not use the input $x$;
  such expressions, without $x$,
  are then split into a reduction function $s(x)$,
  based on the reduction template,
  and a reconstruction function $t(x)$
  based on the rest of the expression.
For example, the $-\F{flip}(0)$ discovered for $\sin(x)$
  is then split into $s(x) = 0 - x$ and $t(y) = -y$,
  and terms $\F{left}(0 - x, \F{?}, -y)$
  and  $\F{right}(0 - x, \F{?}, -y)$ are proposed.
  
Type-directed synthesis has several limitations.
It can only generate
  \F{left}, \F{right}, \F{periodic}, and \F{logarithmic} terms;
  even adding also \F{approx} and \F{polynomial} using Sollya,
  there are still many \name constructs that it cannot synthesize.
The set of reduction templates is also limited,
  and does not include some more-complex range reductions
  used in some state-of-the-art math function implementations.
Finally, type-directed synthesis also depends
  on the mathematical rewrites provided to the e-graph library.
\name's rewrite rules are based
  on the rewrite database of Herbie~\cite{herbie},
  with unsound rules audited by hand and removed.
These rewrite rules include identities
  for core functions such as $\sin(x)$ and $\log(x)$,
  so synthesizing a \F{sin} or \F{log} implementation
  using type-directed synthesis is more an exercise in recall than synthesis.
However, \F{tds} shines when synthesizing implementations
  of compound operations such as $\sin(\pi x)$, $1 - \cos(x)$, or $\sin(x) - x$.
These operations are common in real-world code
  and do not achieve the best possible speed and accuracy
  when implemented naively.
Type-directed synthesis thus provides users
  with get acceptable implementations
  without relying on user expertise.

\subsection{Making Type-Directed Synthesis Practical}

Once the e-graph is grown,
  it must be searched for the expressions of interest,
  specifically expressions
  equivalent to the starting point $f(x)$
  and that do not contain the variable $x$,
  meaning that all calls to $f(x)$ have been replaced
  with reduction templates.
However, the very nature of e-graphs implies
  that the e-graph likely stores a huge number of such terms,
  possibly exponentially or even infinitely large.
To extract a large but potentially useful set of rewrites,
  \name uses \textit{node extraction},
  where one representative expression is extracted from the e-graph
  for each ``e-node'' in the ``e-class'' corresponding to $f(x)$.
In effect, this extracts the largest set of expressions
  where any two expression in the set
  either differ at the root operator
  (such as $0 - \F{flip}(0)$ and $0 + \F{flip}(1)$)
  or have the arguments to the root operator that are non-equivalent
  (such as $0 + \F{flip}(0)$ and $1 + \F{flip}(1)$).
The extraction procedure is configured
  to heavily penalize terms with an $x$ variable,
  and extracted terms that use that variable are discarded.

Node extraction is fast and typically results
  in dozens to hundreds of extracted expressions.
However, far from all of the extracted expressions are useful.
Some restate trivial identities;
  for example, the term $\F{shift}(0)$ stands for
  to expression $f(x + 0)$,
  meaning that it represents the identity $f(x) = f(x + 0)$.
Others restate each other, such as the terms
  $\F{flip}(0)$ and $1 \cdot \F{flip}(0)$.
In these terms have different root operators,
  meaning all of these duplicates will be extracted.
To filter out these duplicates,
  we construct a second e-graph,
  insert all of the extracted terms,
  and again grow that e-graph.
However, this e-graph is not provided with the definition of $\F{thefunc}(x)$.
In effect, this procedure asks the e-graph to identify
  all extracted terms that would be equivalent
  for \emph{any} function $f(x)$,
  and which therefore cannot provide a useful guide
  toward implementing $f(x)$ in particular.
This cuts the number of generated identities substantially,
  typically to only a handful.
Type-directed synthesis can then rapidly refer
  to just this small handful of possible identities
  and rapidly enumerate possible implementations.

Discovering identities such as
  $\sin(x) \leadsto -\F{flip}(0)$ by
  growing an e-graph can take a long time---%
  typically tens of seconds.
It is therefore important that the e-graph 
  is grown as few times as possible.
But since type-directed synthesis adds terms
  that themselves contain holes,
  type-directed synthesis can be invoked repeatedly.
\name's type-directed synthesis engine thus
  grows the e-graph when filling the first hole,
  and then caches the resulting identities.

One more challenge arises
  with \F{periodic} and \F{logarithmic} terms.
These are generated from \F{shift} and \F{scale} templates,
  via terms like $\sin(x) \leadsto -\F{shift}(\pi)$
  and $\log(x) \leadsto \F{scale}(2) + \log(2)$.
The issue is that while these terms contain $t(y)$,
  the \F{periodic} and \F{logarithmic} terms
  require $t(y, k) = t^k(y)$.
This requires solving an inductive equation:
  $t(y, 0) = y$ and $t(y, n + 1) = t(t(y, n))$.
To solve this inductive equation,
  \name uses e-graph intersection~\cite{egraph-intersect}.
Specifically, \name constructs and grows
  a series of e-graphs $E_1, E_2, \dotsc$
  where e-graph $E_k$ contains:
\begin{itemize}
\item The term $t(y, k)$ and equality $t(y, k) = t(t(\dotsb(y)))$
\item The variable $n$ and equality $n = k$
\end{itemize}
This series of e-graphs is then \emph{intersected},
  resulting in a new e-graph $E^*$
  that only contains terms and equalities that were true
  in each intersected e-graph $E_k$.
For example, consider $t(y) = y + \log(2)$,
  which is needed to synthesize the \F{logarithmic}
  reduction for $\log(x)$.
Here, the e-graph $E_1$ initially contains
  the equalities $t(y, 1) = y + \log(2)$ and $n = 1$.
After growing this e-graph,
  it will also contain the equality $t(y, n) = y + n \cdot \log(2)$.
The e-graph $E_2$ initially contains
  the equalities $t(y, 2) = (y + \log(2)) + \log(2)$ and $n = 2$,
  but after growing,
  it too contains $t(y, n) = y + n \cdot \log(2)$.
E-graph intersection can thus,
  in some cases, solve the required inductive equation.
Luckily, the inductive equations required are typically simple,
  and this synthesis method is often successful,
  requiring only a few e-graphs $E_k$ grown for only a few iterations.

\section{Evaluation}
\label{sec:eval}


Our evaluation aims to answer three research questions about \name:
\begin{enumerate}
\item[{\bf RQ1}] Can \name implement state-of-the-art math libraries
  safely, modularly, and tunably?
\item[{\bf RQ2}] Is \name able to tune math libraries for greater speed and accuracy?
\item[{\bf RQ3}] Is \name able to synthesize math libraries automatically?
\end{enumerate}
To this end, we perform three tests,
  first recreating an array of existing math libraries,
  then tuning those libraries for greater speed and accuracy,
  and finally synthesizing implementations of related functions from scratch.
All experiments were performed on an 2020 Apple MacBook Pro
  with an M1 processor, 16GB of RAM,
  Python 3.11.3, Sympy 1.12, Sollya 8.0,
  and Clang 14.0.3 run with {\tt -O3 -mtune=native -DNDEBUG}.
Error and runtime measurements are performed
  using \name's native tools, as described in \Cref{sec:measure}.

\subsection{Re-creating existing libraries (RQ1)}
\label{sec:recreation}

Existing math libraries are time-tested,
  with both algorithms and tuning parameters persisting for decades.
To determine whether \name can express the techniques required
  in state-of-the-art implementations,
  we chose 8 implementations from three widely-used libraries
  (Sun's fdlibm, perhaps the most widely copied math library;
   VDT, a modern variant of the well-known Cephes library;
   and AMD's Optimizing CPU Libraries)
  and attempted to reimplement each in \name.
The libraries, functions, and implementations
  are listed in \Cref{fig:eval1},
  along with the speed and accuracy
  of both the original implementation and \name's reimplementation.
For each function,
  we started by reading code comments,
  which typically had an English-language description
  of the high-level algorithm behind the implementation.
We then implemented that algorithm in \name
  and tweaked the tuning parameters
  until we could match the original implementation's accuracy and speed.
When the original library used techniques
  not currently implemented in \name,
  or when we found flaws in the original library,
  we prioritized keeping to the spirit of the original implementation
  over accuracy or speed.
\name's reimplementations
  are roughly the same speed and accuracy,
  averaging 26\% more accurate
  but 14\% slower.
Most of that slow-down is from two related implementations;
  if those are excluded, \name's reimplementations
  are 0.5\% slower on average.
Moreover, \name's reimplementations
  are much shorter than the originals,
  as shown by the line of code counts in \Cref{fig:eval1}.
These line of code counts understate \name's advantage,
  because the C code of the original libraries
  often uses complex features
  such as pointers, unions, casts, and macros,
  visible in \Cref{fig:fdlibmlog},
  while \name's DSL does not.%
\footnote{Of course \name's generated C code
  does use these features because
  they are necessary for maximum performance.}
Many of \name's lines of code, by contrast,
  are simply coefficients for \F{polynomial} terms.
To get a better sense of \name's flexibility
  and how it can match state-of-the-art implementations,
  we describe the case studies in \Cref{fig:eval1}
  in more detail.

\begin{figure}
  \begin{tabular}{|l|l|l|rr|rr|rr|} \hline
            &                      &             & \multicolumn{2}{c|}{Error}   & \multicolumn{2}{c|}{Runtime} & \multicolumn{2}{c|}{Source Lines} \\
    Library & Func                 & Domain      & Original     & MLM           & Original  & MLM        & Original & MLM    \\ \hline
    AOCL    & \texttt{fast\_asin}  & [0, 0.5]    & \bf 7.50e-17 & \bf 7.50e-17  & 7.34      & \bf 7.26   & 56       & \bf 24 \\
            &                      & [-1, 1]     & \bf 2.69e-16 & 2.70e-16      & \bf 17.93 & 18.47      &          &        \\
            & \texttt{fast\_asinf} & [0, 0.5]    & \bf 3.63e-8  & \bf 3.63e-8   & 7.41      & \bf 7.33   & 27       & \bf 14 \\
            &                      & [-1, 1]     & \bf 1.60e-7  & \bf 1.60e-7   & 19.02     & \bf 18.78  &          &        \\
            & \texttt{opt\_asinf}  & [0, 0.5]    & \bf 3.78e-8  & \bf 3.78e-8   & 7.32      & \bf 7.27   & 40       & \bf 18 \\
            &                      & [-1, 1]     & \bf 5.96e-8  & \bf 5.96e-8   & 18.10     & \bf 17.62  &          &        \\
            & \texttt{ref\_asin}   & [0, 0.5]    & 6.49e-17     & \bf 6.40e-17  & 7.24      & \bf 12.51  & 41       & \bf 21 \\
            &                      & [-1, 1]     & 1.66e-16     & \bf 1.44e-16  & \bf 17.22 & 25.43      &          &        \\
    \hline
    fdlibm  & \texttt{asin}        & [0, 0.5]    & \bf 6.02e-17 & 6.03e-17      & \bf 7.20  & 11.22      & 52       & \bf 25 \\
            &                      & [-1, 1]     & 1.66e-16     & \bf 1.44e-16  & \bf 16.37 & 25.62      &          &        \\
            & \texttt{log}         & [1.7, 2.4]  & 7.43e-17     & \bf 7.04e-17  & 7.44      & \bf 7.31   & 61       & \bf 23 \\
            &                      & [1, 50]     & \bf 2.64e-16 & 2.76e-16      & \bf 7.04  & 8.24       &          &        \\
    \hline
    VDT     & \texttt{cos}         & [0, 0.7]    & \bf 1.34e-16 & 1.38e-16      & 7.46      & \bf 7.29   & 50       & \bf 27  \\
            &                      & [-4, 4]     & \bf 1.42e-16 & 1.44e-16      & 7.17      & \bf 7.05   &          &         \\
            & \texttt{exp}         & [-0.3, 0.3] & \bf 2.69e-16 & \bf 2.69e-16  & 7.39      & \bf 7.27   & 30       & \bf 16  \\
            &                      & [-20, 20]   & \bf 6.40e-8  & \bf 6.40e-8   & \bf 7.04  & 7.05       &          &         \\
            &                      & [0, 50]     & \bf 6.77e5   & \bf 6.77e5    & 7.55      & \bf 7.55   &          &         \\
    \hline
    \end{tabular}

        \caption{Maximum absolute error and average runtime for each of the
            implementations reimplemented using the \name DSL.
            \name's variation reached the exact same output for tested point in
            VDT's $\exp(x)$ and two of AOCL's $\sin^{-1}(x)$ variations.
            Other implementations have very close error values.
            Performance is matched across most implementations with AOCL's
            reference and fdlibm's versions of $\sin^{-1}(x)$ being slower
            when reimplemented in \name.}
        \label{fig:eval1}
        \Description{\todo{desc}}
\end{figure}

\paragraph{fdlibm's \F{log}}

Sun's fdlibm library uses a complex and clever
  implementation of the simple natural logarithm \F{log}.
Like most implementations,
  it begins by first extracting and saving the exponent bits,
  reducing the input range to $[2^{-1/2}, 2^{1/2}]$
  (implemented with \name's \F{logarithmic} operator)
  and then subtracting 1 from the input
  (implemented with \name's composition operator),
  which is exact, thanks to Sterbenz's lemma~\cite{sterbenz}.
That leaves the task of implementing $\log(f + 1)$
  on a range of approximately $[-0.3, 0.4]$,
  where $f = x - 1$.
Next, fdlibm performs the substitution $s = f / (2 + f)$,
  transforming the target function into $\log(1 + s) - \log(1 - s)$
  on the range $[-0.18, 0.17]$;
  this is again represented in \name with a composition.
The advantage of this transformation
  is that the new target function is odd,
  meaning that it can be approximated by a polynomial
  with only odd powers of $x$.
Fdlibm then uses a polynomial approximation,
  organized in a somewhat strange way
  with the terms split modulo 4 and the leading term split out;
  in \name, this strange organization can be replicated
  with addition terms, though perhaps it could also be represented
  by a new value for the \F{method} tuning parameter
  for the \F{polynomial} term.
Finally, to combat rounding error in the computation of $s$,
  fdlibm uses a rewrite of $2s \leadsto f - s \cdot f$,
  iterated once for most of the range and twice
  for a particularly difficult interval from $0.38$ to $0.41$.
This rewrite is expressed in \name with the \F{rewrite} operator,
  which has the advantage of verifying
  that the rewrite is mathematically correct.
Finally, the reconstruction operation
  is implemented in the standard way,
  with the original exponent bits multiplied by $\log(2)$
  and added to the final value.
\name's reimplementation of fdlibm's \F{log}
  is shown in \Cref{fig:mlmlog},
  with tuning parameters hidden to highlight the overall algorithm.

One technique in fdlibm's \F{log} implementation,
  however, could not be replicated in \name.
In fdlibm's \F{log},
  the reconstruction operation $y + k \cdot \log(2)$
  and the polynomial summation $2s + s \cdot R$
  are interleaved, with terms summed from smallest to largest.
This reduces compounding rounding error,
  but interferes with \name's ``black-box'' approach to compilation.
To approximate the effects of reordering,
  our \name implementation uses double-double arithmetic~\cite{dekker},
  which achieves similar accuracy to fdlibm
  but at the cost of additional floating-point operations.
In the future, we hope to add support for this technique
  by adding an uninterpreted-sum data type to \name.
Adding uninterpreted sums would simply concatenate compile-time lists of terms,
  and converting an uninterpreted sum into a floating-point value
  would perform interval analysis to sort terms by magnitude
  before adding them together.
Preliminary tests suggest that this would allow
  \name to match fdlibm's \F{log} exactly.

\begin{figure}
    \begin{code}
    \F{R} = \F{polynomial}([4:\ldots]) + \F{polynomial}([2:\ldots]) \\
    \F{core} = \F{approx}(\log(1+s) - \log(1-s), [-0.17, 0.17], \sci{5}{-19},
    2\cdot s + s\cdot R), \\
    \F{logarithmic}(2, \\
    \> (f = x-1) ; \F{split}([ \\
    \>\> [\sci{-9.5}{-7}, \sci{9.5}{-7}] \mapsto \\
    \>\>\> \F{approx}(\log(f+1), [\sci{-9.5}{-7}, \sci{9.5}{-7}], \sci{3}{-25}, \\
    \>\>\>\> \F{polynomial}([1:1, 2:-0.5, 3:0.33])), \\
    \>\> [-0.29, 0.38] \mapsto \\
    \>\>\> (s = f/(2+f)) ; \F{rewrite}(\F{core}, \\
    \>\>\>\> (2\cdot s) \mapsto \F{fma}(-s, f, f)), \\
    \>\> [0.38, 0.41] \mapsto \\
    \>\>\> (s = f/(2+f)) ; \F{rewrite}(\F{core}, \\
    \>\>\>\> (2\cdot s) \mapsto (f - f\cdot f \cdot 0.5 + (s\cdot f \cdot f\cdot 0.5)))]), \\
    \> y + k\cdot \log(2))
    \end{code}

    \caption{A implementation of fdlibm's $\log(x)$ function in \name.
      Polynomial coefficients and tuning parameters are omitted for clarity,
      and various numeric constants are shortened.}
    \label{fig:mlmlog}
    \Description{A \name implementation of the logarithm function
      leveraging the key ideas behind fdlibm's implementation of the same.}
\end{figure}

\paragraph{VDT's \F{cos}}
Like many cosine implementations,
  VDT reduces the input range to $[0, \frac\pi4]$
  using a Cody-Waite additive range reduction
  (implemented using \name's \F{periodic} operator)
  and reconstructing the result
  using the count $k$ of the reduction modulo 4.
On the reduced range,
  VDT uses a polynomial approximation.
Most of the terms of this polynomial approximation
  are generated by Sollya's \F{fpminimax} algorithm.
However, the first two terms of this polynomial
 are fixed to $1$ and $-\frac12$,
 which are the first two terms of the Taylor expansion for cosine;
 this common practice guarantees higher accuracy near $0$.
Moreover, instead of using the standard Horner form,
  $a_0 + x \cdot (a_1 + x \cdot (a_2 + \dotsb))$,
  VDT computes the first two terms of the polynomial separately,
  as \texttt{1.0 - zz * .5 + zz * zz * get\_cos\_px(zz)},
  where \texttt{zz} is equal to $x^2$
  and where \texttt{get\_cos\_px} implements
  the rest of the polynomial.

\begin{figure}
    \includegraphics[width=\linewidth]{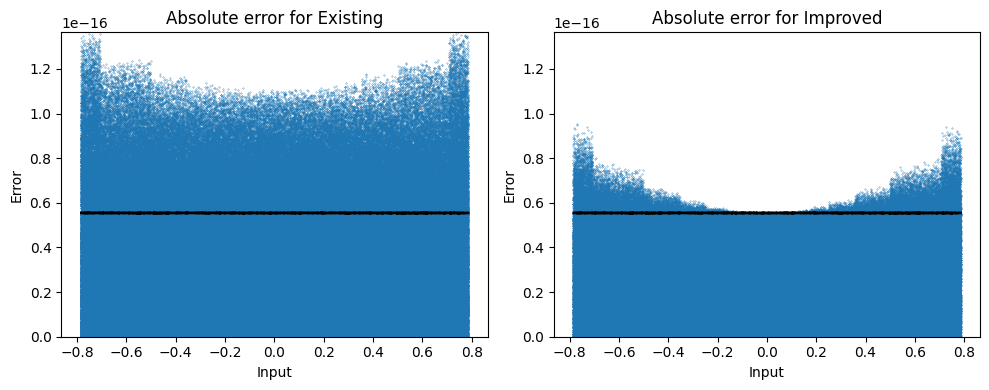}

    \hfill
    \begin{code}
        \F{polynomial}([0: 1, 2: -0.5]) + {}\\
        \F{approx}(\cos(x) - 1 + \frac{x^2}2, [-\frac{\pi}{4}, \frac{\pi}{4}], \sci{6}{-20}, \\
        \>\F{polynomial}([\dotsc]))
    \end{code}%
    \hfill%
    \begin{code}
        \F{approx}(\cos(x), [-\frac{\pi}{4}, \frac{\pi}{4}], \sci{6}{-20}, \\
        \> \F{polynomial}([
            0: 1, 2:  -0.5,
            \dotsc
        ]), \F{split}{=}1)
    \end{code}
    \hfill

    \caption{Absolute error graphs and \name code
      for the core of the VDT \F{cos} implementation.
      On the left, our attempt to replicate
      VDT's actual implementation;
      on the right, a more natural \name implementation
      that fixes a seeming bug in VDT's \F{cos}.
    The bug fails to parenthesize \texttt{1.0 - zz * .5} in
      \texttt{1.0 - zz * .5 + zz * zz * get_cos_px(zz)}.}
    \label{fig:vdtcos}
    \Description{\todo{desc}}
\end{figure}

Splitting out the first term of the polynomial like this
  can reduce rounding error,
  and \name offers it via the \F{split} tuning parameter
  for \F{polynomial} terms.
However, seemingly as the result of an oversight,
  the VDT implementation adds the terms
  largest-to-smallest instead of smallest-to-largest.%
  \footnote{We have reported this bug to the VDT developers.}
In C, this is an easy mistake to make,
  since it depends on the associativity of the \F{-} and \F{+} operations.
However, it dramatically increases the error (see \Cref{fig:vdtcos}),
  with seemingly no benefits.%
\footnote{Changing the order of summation
  does shorten the arithmetic critical path,
  so could hypothetically lead to speed ups,
  but we were unable to measure this effect;
  in any case, our \name reimplementation is faster.}
\name's \F{split} tuning parameter does not have this bug,
  and achieves better accuracy;
  to replicate the actual VDT implementation for \Cref{fig:eval1}
  requires complex contortions (see \Cref{fig:vdtcos})
  and is still not exactly identical to VDT.

\paragraph{AOCL's many \F{asin}s}

The AMD Optimizing CPU Libraries include
  not just one but four \F{asin} implementations:
  \F{ref} and \F{opt} variants with greater accuracy,
  and \F{fast} variants for lower accuracy but greater speed.
We re-implemented all four in \name;
  all four share similar high-level algorithms,
  though the \F{fast} variants use fewer polynomial terms
  and the \F{ref} variant uses a modified Pad\'e approximation
  instead of a polynomial approximation.
Some variants even share polynomial coefficients.
We also implemented the very similar fdlibm \F{asin} function.
However, the five implementations are tuned differently:
  the \F{fast\_asin} variant, for example, adds a \F{split} term,
  while the \F{opt\_asinf} variant
  does most of the computation in double precision
  before casting to a single-precision result.

There is one aspect of these implementations
  that we were unable to replicate in \name.
All four implementations use the identity
\(
\frac\pi2 - 2 \cdot \F{asin}\left(\sqrt{(1 - x)/2}\right).
\)
  for range reduction and reconstruction.
\name's \F{right} operator ably expresses this reduction;
  however, the core polynomial approximation used for \F{asin}
  uses only even powers of the reduced value $x$.
On range-reduced inputs,
  that would normally mean computing $\sqrt{\ldots}^2$,
  an expression that introduces unnecessary error.
The AOCL implementations
  save the $(1 - x) / 2$ value before performing the square root
  to avoid this.
\name's \F{rewrite} operator
  cannot be used to similar effect
  because it operates on one IR node at a time,
  whereas the square root and the square operation
  are located in different nodes.
In the future we hope to add
  cross-block matching to \F{rewrite},
  allowing it to express this trick.

Unfortunately,
  our reimplementations of AOCL's \F{ref\_asin} function,
  and the similar fdlibm \F{asin} function,
  are dramatically slower than the original implementation.
The slow-down is due to a fairly simple cause:
  the \name implementation compiles the whole reduction operation
  $\sqrt{(1 - x)/2}$ to double-double precision,
  but in fact the subtraction and division operations are exact
  and can be implemented with standard double precision.
While it is possible to address this issue in \name,
  it would require changes to the implementation
  beyond simple tweaking of tuning parameters,
  and we felt such changes went against the spirit of \name.
In the future, we hope to detect exact operations like this
  using interval analysis, leveraging the intervals
  available in \name's type system,
  similar to \citet{verified-math-h}.

\subsection{Tuning Implementations for Speed and Accuracy (RQ2)}
\label{sec:improvement}

\begin{figure}
    \begin{tabular}{|l|l|l|rr|rr|r|} \hline
                &                    &               & \multicolumn{2}{c|}{Error}  & \multicolumn{2}{c|}{Runtime} & \\
        Library & Func               & Domain        & Original     & MLM          & Original  & MLM      & + \\ \hline
        AOCL     & \F{fast\_asin\_better} & [0.0, 0.5]    & 7.50e-17     & \bf 6.35e-17 & 7.27      & \bf 7.20 &  2 \\
                &                    & [-1.0, 1.0]   & \bf 2.69e-16 & 2.83e-16     & \bf 17.96 & 18.52    &  \\
        \hline
        VDT     & \F{cos\_better}        & [0.0, 0.78]   & 1.34e-16     & \bf 9.53e-17 & 7.37      & \bf7.19  &  0 \\
                &                    & [-4.0, 4.0]   & 1.42e-16     & \bf 1.12e-16 & 7.44      & \bf7.29  &  \\
                & \F{cos\_faster}        & [0.0, 0.78]   & 1.34e-16     & \bf 1.13e-16 & 7.43      & \bf7.25  &  4 \\
                &                    & [-4.0, 4.0]   & 1.42e-16     & \bf 1.12e-16 & 7.17      & \bf7.05  &  \\
                & \F{exp\_better}        & [-0.34, 0.34] & 2.69e-16     & \bf 1.54e-16 & 7.29      & \bf7.23  &  2 \\
                &                    & [-20.0, 20.0] & 6.40e-8      & \bf 4.45e-8  & \bf 7.04  & \bf7.04  &  \\
                &                    & [0.0, 50.0]   & 6.77e5       & \bf 5.81e5   & 7.56      & \bf7.55  &  \\
                & \F{exp\_faster}        & [-0.34, 0.34] & 2.69e-16     & \bf 1.53e-16 & 7.38      & \bf7.25  &  2 \\
                &                    & [-20.0, 20.0] & 6.40e-8      & \bf 4.96e-8  & \bf 7.04  & \bf7.04  &  \\
                &                    & [0.0, 50.0]   & 6.77e5       & \bf 5.81e5   & 7.54      & \bf7.54  &  \\
        \hline
        \end{tabular}

        \caption{Variations on some of the reimplemented library functions.
        The ``better'' variations were made to decrease error, while the
        ``faster'' versions were made to increase speed.
        The ``+'' column counts lines added in the variation.
        This column does not subtract lines removed;
          typically, more lines are added than removed,
          because polynomial coefficients are re-synthesized.
        }
        \label{fig:eval2}
        \Description{\todo{desc}}
\end{figure}

To test that \name makes it easy to tune math library implementations,
  we construct 5 variations of the implementations in \Cref{fig:eval1}
  that use simple, widely applicable techniques
  to improve speed or accuracy.
These variations change only a few lines,
  and never add more lines than they remove,
  and almost always result in implementations
  that are both faster and more accurate.

To improve the accuracy of AOCL's \F{fast\_asin},
  we resynthesize the core function approximation
  using Sollya's \F{fpminimax} algorithm via \name.
AOCL's \F{fast\_asin} actually already used
  coefficients synthesized in this way;
  however, instead of being directly synthesized for \F{fast\_asin},
  they were just the coefficients from \F{opt\_asinf}
  but truncated to form a shorter polynomial.
This is not quite as accurate as resynthesizing the polynomial.
The result was somewhat more accurate over the core $[0, 0.5]$ domain,
  but actually less accurate over the whole domain $[-1, 1]$.
The resynthesized polynomial could still be useful
  (if, say, the user knows their inputs are in $[-0.5, 0.5]$)
  but it also demonstrates an important aspect of \name's design:
  floating point error is fundamentally not composable,
  so tuning and experimentation is often the best path.
\name therefore focuses on making it easy
  to measure, tune, and test floating-point speed and accuracy.

We made a more substantial change to VDT's \F{exp}.
The original implementation used an unusual rational polynomial
  $1 + 2 \frac{p}{q - p}$,
  where $p$ has three terms and $q$ has four.
We synthesized polynomial approximations using \F{remez}%
\footnote{While \F{fpminimax} is typically more accurate,
  \F{remez} is much faster so a better match for testing many sizes.}
  with between 7 and 12 terms, and found
  that an 11-term polynomial approximation had comparable speed
  to the existing approximation.
Note that, while this polynomial has more terms,
  it avoids an expensive division operation,
  meaning more terms can be used.
The synthesized polynomial therefore has comparable speed
  while improving error substantially.
We also tested a faster version,
  which changes the polynomial evaluation scheme from Horner to Estrin.
We expected this to produce a speedup,
  because 11-term polynomials are long enough
  for critical path length to become important.
However, we were unable to measure any difference,
  again underscoring the importance of measurement.

Finally, noting the seeming oversight
  in VDT's \F{cos} implementation,
  we implemented a ``fixed'' version of VDT \F{cos},
  which is substantially more accurate than the original.
We also experimented with a faster version,
  which resynthesizes the polynomial coefficients,
  uses Estrin polynomial evaluation,
  and splits out one term instead of two.
This was slightly faster on the full range,
  likely the result of Estrin's shorter critical path,
  but less accurate on the narrow range.
This again demonstrates the benefit
  of \name's empirical, interactive design:
  we do not know of a systematic way
  to estimate the costs and benefits of Estrin over Horner form,
  but testing both options made the trade-offs clear.

\subsection{Automated Synthesis of New Libraries (RQ3)}
\label{sec:creation}

\begin{figure}
  \begin{tabular}{|l|l|rr|rr|} \hline
                   &               & \multicolumn{2}{c|}{Error}     & \multicolumn{2}{c|}{Runtime} \\
     Func          & Domain        & Lib            & MLM           & Lib & MLM  \\ \hline
     $1-\cos(x)$   & [-1.0, 1.0]   & 9.12e-17       & \bf{5.78e-17} & 12.17      & \bf 7.50 \\
                   & [-2.0, 2.0]   & \bf{1.41e-16}  & 2.91e-16      & 15.87      & \bf 7.49 \\
                   & [-8.0, 8.0]   & \bf{1.97e-16}  & 8.72e-16      & 19.11      & \bf 7.48 \\
                   & [-32.0, 32.0] & \bf 2.04e-16   & 1.39e-15      & 19.08      & \bf 7.40 \\
                   \hline
     $\sin(x) - x$ & [-1.0, 1.0]   & 9.31e-17       & \bf{8.79e-17} & 12.29      & \bf 7.54 \\
                   & [-2.0, 2.0]   & \bf{1.75e-16}  & 3.63e-16      & 16.36      & \bf 7.51 \\
                   & [-8.0, 8.0]   & \bf{5.31e-16}  & 9.70e-15      & 18.99      & \bf 7.61 \\
                   & [-32.0, 32.0] & \bf 1.86e-15   & 1.18e-14      & 18.63      & \bf 7.40 \\
                  \hline
     $\sin(\pi x)$ & [-1.0, 1.0]   & \bf 3.46e-16   & 1.10e-15      & 22.36      & \bf 8.16 \\
                   & [-2.0, 2.0]   & \bf 6.85e-16   & 1.03e-15      & 19.82      & \bf 7.49 \\
                   & [-8.0, 8.0]   & 2.73e-15       & \bf 1.12e-15  & 19.52      & \bf 7.50 \\
                   & [-32.0, 32.0] & 1.08e-14       & \bf 1.05e-15  & 19.43      & \bf 7.51 \\
    \hline
  \end{tabular}

  \caption{Fully synthesized implementations of compound math functions
         and their straightforward C counterparts.
        All "lib" implementations are using the system libm.}
        \label{fig:eval3}
        \Description{\todo{desc}}
\end{figure}

To test the \name's combined
  type-directed and Sollya-based synthesis,
  we constructed implementations of $1-\cos(x)$, $\sin(x) - x$, and $\sin(\pi x)$
  entirely from scratch,
  and compared them to the naive implementation of these functions
  where, for example, $\sin(\pi x)$ is implemented as \texttt{sin(M_PI * x)},
  with \F{sin} referencing the GLibC standard library.
To synthesize each implementation,
  we constructed a hole of the appropriate type and called \F{synthesize}.
Polynomials were set to be 14 terms and created using the \F{remez} algorithm.
No further tuning was performed;
  when \name produced multiple expressions, the most accurate was chosen.
Results are shown in \Cref{fig:eval3}.

All synthesized implementations are significantly faster
  than the naive library-based implementation.
For $1 - \cos(x)$ and $\sin(x) - x$, the \name-synthesized implementations
  are more accurate on narrow domains,
  but become less accurate as the domain widens;
  this is because \name-synthesized implementations
  by default use the \F{naive} method for \F{periodic} terms
  instead of the more-accurate \F{cody-waite} method.
For $\sin(\pi x)$, by contrast, the \name-synthesized implementation
  shows a constant accuracy as the domain widens,
  while the naive library-based implementation
  grows progressively less accurate
  as the domain grows wider.
This is because in the naive library-based implementation,
  multiplying by $\pi$ and then reducing modulo $2\pi$ introduces error,
  while the \name-synthesized implementation
  can perform a fast \F{periodic} reduction by $2$, which is exact.
Each of these implementations can be further tuned by the user,
  including tuning the \F{periodic} reduction method,
  introducing higher accuracy, 
  or switching to \F{fpminimax} polynomial approximations.
While \name's synthesized methods cannot match
  hand-tuned expert-written functions,
  they provide an alternative for users without the expertise
  to write their own.

\section{Related Work}
\label{sec:related}

Mathematical function implementation, at its core,
  replaces a hard-to-compute function
  and with simpler operations that are easily computable.
Work in this area starts before electric computers
  with the work of people like
  Taylor~\cite{taylor}, Chebyshev~\cite{chebyshev}, and Remez~\cite{remez},
  who created the foundations of approximation theory.
This work gained practical importance
  with the advent of vacuum tube computers~\cite{cecil}.
As computers increased in speed,
  control over accuracy became important, with important contributions
  by \citet{dekker}, \citet{kahan, kahan-log}, and \citet{higham}.
The roots of modern math libraries go back to Cody and Waite's
  "Software Manual for the Elementary Functions"~\citep{cody-waite},
  which provided a detailed guide for many common functions.
Modern references on math function function implementation
  include the ``Mathematical-Function Computation Handbook''~\citep{beebe}
  and ``Elementary Functions''~\citep{muller}.

Many extant math libraries trace coefficients and code
  to a couple of especially important historic libraries,
  chief among them Sun's fdlibm library~\cite{fdlibm}.
Other widely-used libraries include
  GNU's GLibC math functions~\cite{glibc},
  the OpenLibm project~\cite{openlibm},
  Cephes~\cite{cephes}, VDT~\cite{vdt},
  Intel's MKL~\cite{mkl}, and AMD's AOCL~\cite{amdlibm}.
All of these libraries achieve accuracy of about 1 ULP of error
  for most functions~\cite{glibc-accuracy,zimmermann},
Four functions from Intel's MKL library
  have had their accuracy verified
  with semi-automated methods~\cite{verified-math-h},
  though it is unclear whether this approach
  scales to other libraries or functions.
A variety of tools support the development of these libraries.
Sollya~\cite{sollya} provides several polynomial approximation algorithms,
  and the Metalibm Python library~\cite{metalibm}
  provides a programmatic interface to it.
Specialized math software such as Magma~\cite{magma},
  Maple~\cite{maple}, Mathematica~\cite{mathematica},
  Matlab~\cite{matlab} (via the Chebfun package~\cite{chebfun}),
  and Sage Math~\cite{sagemath}
  also provide polynomial or rational function approximation packages.
Other tools, such as Flopoco~\cite{flopoco},
  are specialized to hardware implementation.
None of these tools, however,
  provide a high-level language for math function implementation
  that provides safety, modularity, and tunability.

More recently, new research has suggested paths
  toward \emph{correctly rounded} implementations,
  which achieve the lowest possible error of $\frac12$ ULP on all inputs.
The CRLibm library~\cite{crlibm} is one product of this line of work,
  as are the the RLibm project~\cite{rlibm1,rlibm2,rlibm3,jay-p-lim}
  and the in-progress CORE-Math libraries~\cite{core-math}.
The MPFR library also offers correctly rounding implementations
  of the core functions at arbitrary precision~\cite{mpfr},
  though at high runtime cost,
  and libraries that track their precision, such as Arb~\cite{arb},
  can in some cases play a similar role.
Recent advances allows modern correctly rounded libraries
  to match the performance of classic libraries such as GLibC or fdlibm
  while also providing the ultimate accuracy guarantee~\cite{llvmlibm}.
All of these implementations still require
  careful design of range reduction and reconstruction algorithms,
  and could potentially be written in \name.

Even an accurate library function does not guarantee accurate user code.
Many researchers have thus attempted to develop tools
  to bound the worst-case error
  from a given computation~\cite{daisy,fptaylor,satire,real2float}
  or even to discover specific high-error inputs~\cite{s3fp,fpgen}.
Those high-error inputs could then be
  debugged~\cite{fpdebug,herbgrind,positdebug}
  and the code rewritten to improve accuracy.
The Herbie tool~\cite{herbie} attempts
  to automatically rewrite math expressions for higher accuracy
  and, in recent version, greater speed~\cite{pherbie}.
Other packages, such as Salsa~\cite{salsa}, Daisy~\cite{daisy},
  Precimonious~\cite{precimonious}, FPTuner~\cite{fptuner},
  HiFPTuner~\cite{hifptuner}, and POP~\cite{pop}
  provide automated precision adjustment.
Some authors have proposed tools to select~\cite{optuner}
  or synthesize~\cite{daisy-libm}
  the optimal function implementation for a particular call site.
These tools must be used in concert with \name
  to achieve the desired accuracy and speed for application code.

\section{Conclusion}
\label{sec:conclusion}

Implementing math functions has long been challenging
  because the programming languages necessary for high performance
  did not provide safety, modularity, or tunability.
This implied a top-down, monolithic, one-shot implementation style
  that demanded experience and expertise.
\name addresses this with a high-level DSL
  where math library implementations can be expressed, tuned,
  and then compiled to highly accurate, highly performant code.
Implementations in \name are automatically checked for
  syntactic and semantic wellformedness,
  and tuning parameters allow low-level control of speed and accuracy
  while remaining separate from high-level algorithm design.
\name thus allows users---%
  even, thanks to synthesis, users without extensive expertise---%
  to interactively write and tune math function implementations.
We used \name to re-implement 8 state-of-the-art math libraries
  with comparable (often better) speed and accuracy,
  make 5 improved versions,
  and synthesize 3 implementations from scratch.

In the future, we hope to extend \name to incorporate
  more techniques used in state-of-the-art math libraries.
The biggest gap in MegaLibm is the use of lookup tables in computations, which
  have become increasingly important.
Support for tabular approximations
  is necessary for new implementations
  such as RLibm~\cite{jay-p-lim} and CORE-Math~\cite{core-math}.
A key design question is how to support mixed methods
  such as tables of polynomial coefficients
  or polynomial interpolation schemes.
RLibm could be integrated
  as a synthesis tool for tabular approximations.
We also hope to integrate algorithms
  such as Payne-Hanek reduction or compensated Horner evaluation.
We would like to support number representations
  such as half-precision, bfloat16~\cite{bfloat16},
  posits~\cite{posits}, and FP8~\cite{fp8},
  and improve support for compound representations
  such as triple-/quadruple-double support.
We'd also like to use interval analysis and uninterpreted sums
  to improve the existing double-double support
Finally, we'd like to add formal verification backends to \name,
  so that users could send \name implementations to
  Gappa~\cite{gappa}, FPTaylor~\cite{fptaylor}, Satire~\cite{satire}, or Daisy~\cite{daisy}.
While full formal verification of a math library implementation
  is likely still beyond the state of the art,
  these tools would be useful during interactive development.

\begin{acks}
We also thank our anonymous reviewers for their guidance and valuable
   suggestions while preparing the final version of this paper.
This work was supported by NSF award 2119939.
This material is based upon work supported by the U.S. Department of Energy,
  Office of Science, Office of Advanced Scientific Computing Research,
  ComPort: Rigorous Testing Methods to Safeguard Software Porting, under
  Award Number 10061193.
\end{acks}

\bibliography{paper}


\begin{thebibliography}{67}


\ifx \showCODEN    \undefined \def \showCODEN     #1{\unskip}     \fi
\ifx \showDOI      \undefined \def \showDOI       #1{#1}\fi
\ifx \showISBNx    \undefined \def \showISBNx     #1{\unskip}     \fi
\ifx \showISBNxiii \undefined \def \showISBNxiii  #1{\unskip}     \fi
\ifx \showISSN     \undefined \def \showISSN      #1{\unskip}     \fi
\ifx \showLCCN     \undefined \def \showLCCN      #1{\unskip}     \fi
\ifx \shownote     \undefined \def \shownote      #1{#1}          \fi
\ifx \showarticletitle \undefined \def \showarticletitle #1{#1}   \fi
\ifx \showURL      \undefined \def \showURL       {\relax}        \fi
\providecommand\bibfield[2]{#2}
\providecommand\bibinfo[2]{#2}
\providecommand\natexlab[1]{#1}
\providecommand\showeprint[2][]{arXiv:#2}

\bibitem[Aanjaneya et~al\mbox{.}(2022)]%
        {rlibm2}
\bibfield{author}{\bibinfo{person}{Mridul Aanjaneya}, \bibinfo{person}{Jay~P.
  Lim}, {and} \bibinfo{person}{Santosh Nagarakatte}.}
  \bibinfo{year}{2022}\natexlab{}.
\newblock \showarticletitle{Progressive Polynomial Approximations for Fast
  Correctly Rounded Math Libraries}. In \bibinfo{booktitle}{\emph{Proceedings
  of the 43rd ACM SIGPLAN International Conference on Programming Language
  Design and Implementation}} (San Diego, CA, USA) \emph{(\bibinfo{series}{PLDI
  2022})}. \bibinfo{publisher}{Association for Computing Machinery},
  \bibinfo{address}{New York, NY, USA}, \bibinfo{pages}{552–565}.
\newblock
\showISBNx{9781450392655}
\urldef\tempurl%
\url{https://doi.org/10.1145/3519939.3523447}
\showDOI{\tempurl}


\bibitem[Aanjaneya and Nagarakatte(2023)]%
        {rlibm3}
\bibfield{author}{\bibinfo{person}{Mridul Aanjaneya} {and}
  \bibinfo{person}{Santosh Nagarakatte}.} \bibinfo{year}{2023}\natexlab{}.
\newblock \showarticletitle{Fast Polynomial Evaluation for Correctly Rounded
  Elementary Functions Using the RLIBM Approach}. In
  \bibinfo{booktitle}{\emph{Proceedings of the 21st ACM/IEEE International
  Symposium on Code Generation and Optimization}} (Montr\'{e}al, QC, Canada)
  \emph{(\bibinfo{series}{CGO 2023})}. \bibinfo{publisher}{Association for
  Computing Machinery}, \bibinfo{address}{New York, NY, USA},
  \bibinfo{pages}{95–107}.
\newblock
\showISBNx{9798400701016}
\urldef\tempurl%
\url{https://doi.org/10.1145/3579990.3580022}
\showDOI{\tempurl}


\bibitem[{AMD}(2021)]%
        {amdlibm}
\bibfield{author}{\bibinfo{person}{{AMD}}.} \bibinfo{year}{2021}\natexlab{}.
\newblock \bibinfo{title}{{AMD} Math Library ({LibM})}.
\newblock
\newblock
\urldef\tempurl%
\url{https://developer.amd.com/amd-aocl/amd-math-library-libm/}
\showURL{%
\tempurl}


\bibitem[Bard et~al\mbox{.}(2019)]%
        {real2float}
\bibfield{author}{\bibinfo{person}{Joachim Bard}, \bibinfo{person}{Heiko
  Becker}, {and} \bibinfo{person}{Eva Darulova}.}
  \bibinfo{year}{2019}\natexlab{}.
\newblock \showarticletitle{Formally Verified Roundoff Errors Using SMT-based
  Certificates and Subdivisions}. In \bibinfo{booktitle}{\emph{Formal Methods
  -- The Next 30 Years}}, \bibfield{editor}{\bibinfo{person}{Maurice~H. ter
  Beek}, \bibinfo{person}{Annabelle McIver}, {and} \bibinfo{person}{Jos{\'e}~N.
  Oliveira}} (Eds.). \bibinfo{publisher}{Springer International Publishing},
  \bibinfo{address}{Cham}, \bibinfo{pages}{38--44}.
\newblock
\showISBNx{978-3-030-30942-8}


\bibitem[Beebe(2017)]%
        {beebe}
\bibfield{author}{\bibinfo{person}{Nelson H.~F. Beebe}.}
  \bibinfo{year}{2017}\natexlab{}.
\newblock \showarticletitle{The Mathematical-Function Computation Handbook}. In
  \bibinfo{booktitle}{\emph{Cambridge International Law Journal}}.
\newblock


\bibitem[Behnam and Bojnordi(2020)]%
        {posits}
\bibfield{author}{\bibinfo{person}{Payman Behnam} {and} \bibinfo{person}{Mahdi
  Bojnordi}.} \bibinfo{year}{2020}\natexlab{}.
\newblock \bibinfo{title}{Posit: {A} {Potential} {Replacement} for {IEEE} 754}.
\newblock
\newblock
\urldef\tempurl%
\url{https://www.sigarch.org/posit-a-potential-replacement-for-ieee-754/}
\showURL{%
\tempurl}


\bibitem[{Ben Khalifa} and Martel(2022)]%
        {pop}
\bibfield{author}{\bibinfo{person}{Dorra {Ben Khalifa}} {and}
  \bibinfo{person}{Matthieu Martel}.} \bibinfo{year}{2022}\natexlab{}.
\newblock \showarticletitle{Constrained Precision Tuning}. In
  \bibinfo{booktitle}{\emph{8th International Conference on Control, Decision
  and Information Technologies, CoDIT 2022, Istanbul, Turkey, May 17-20,
  2022}}. \bibinfo{publisher}{{IEEE}}, \bibinfo{pages}{230--236}.
\newblock
\urldef\tempurl%
\url{https://doi.org/10.1109/CoDIT55151.2022.9804011}
\showDOI{\tempurl}


\bibitem[Benz et~al\mbox{.}(2012)]%
        {fpdebug}
\bibfield{author}{\bibinfo{person}{Florian Benz}, \bibinfo{person}{Andreas
  Hildebrandt}, {and} \bibinfo{person}{Sebastian Hack}.}
  \bibinfo{year}{2012}\natexlab{}.
\newblock \showarticletitle{A Dynamic Program Analysis to Find Floating-Point
  Accuracy Problems}.
\newblock \bibinfo{journal}{\emph{SIGPLAN Not.}} \bibinfo{volume}{47},
  \bibinfo{number}{6} (\bibinfo{date}{jun} \bibinfo{year}{2012}),
  \bibinfo{pages}{453--462}.
\newblock
\showISSN{0362-1340}
\urldef\tempurl%
\url{https://doi.org/10.1145/2345156.2254118}
\showDOI{\tempurl}


\bibitem[Boehm(2020)]%
        {api-reals}
\bibfield{author}{\bibinfo{person}{Hans-J. Boehm}.}
  \bibinfo{year}{2020}\natexlab{}.
\newblock \showarticletitle{Towards an API for the Real Numbers}. In
  \bibinfo{booktitle}{\emph{Proceedings of the 41st ACM SIGPLAN Conference on
  Programming Language Design and Implementation}} (London, UK)
  \emph{(\bibinfo{series}{PLDI 2020})}. \bibinfo{publisher}{Association for
  Computing Machinery}, \bibinfo{address}{New York, NY, USA},
  \bibinfo{pages}{562--576}.
\newblock
\showISBNx{9781450376136}
\urldef\tempurl%
\url{https://doi.org/10.1145/3385412.3386037}
\showDOI{\tempurl}


\bibitem[Bosma et~al\mbox{.}(1997)]%
        {magma}
\bibfield{author}{\bibinfo{person}{Wieb Bosma}, \bibinfo{person}{John Cannon},
  {and} \bibinfo{person}{Catherine Playoust}.} \bibinfo{year}{1997}\natexlab{}.
\newblock \showarticletitle{The {M}agma algebra system. {I}. {T}he user
  language}.
\newblock \bibinfo{journal}{\emph{J. Symbolic Comput.}} \bibinfo{volume}{24},
  \bibinfo{number}{3-4} (\bibinfo{year}{1997}), \bibinfo{pages}{235--265}.
\newblock
\showISSN{0747-7171}
\urldef\tempurl%
\url{https://doi.org/10.1006/jsco.1996.0125}
\showDOI{\tempurl}
\newblock
\shownote{Computational algebra and number theory (London, 1993)}.


\bibitem[Briggs and Panchekha(2022)]%
        {optuner}
\bibfield{author}{\bibinfo{person}{Ian Briggs} {and} \bibinfo{person}{Pavel
  Panchekha}.} \bibinfo{year}{2022}\natexlab{}.
\newblock \showarticletitle{Choosing Mathematical Function Implementations for
  Speed and Accuracy}. In \bibinfo{booktitle}{\emph{Proceedings of the 43rd ACM
  SIGPLAN International Conference on Programming Language Design and
  Implementation}} (San Diego, CA, USA) \emph{(\bibinfo{series}{PLDI 2022})}.
  \bibinfo{publisher}{Association for Computing Machinery},
  \bibinfo{address}{New York, NY, USA}, \bibinfo{pages}{522--535}.
\newblock
\showISBNx{9781450392655}
\urldef\tempurl%
\url{https://doi.org/10.1145/3519939.3523452}
\showDOI{\tempurl}


\bibitem[Chiang et~al\mbox{.}(2017)]%
        {fptuner}
\bibfield{author}{\bibinfo{person}{Wei-Fan Chiang}, \bibinfo{person}{Mark
  Baranowski}, \bibinfo{person}{Ian Briggs}, \bibinfo{person}{Alexey Solovyev},
  \bibinfo{person}{Ganesh Gopalakrishnan}, {and} \bibinfo{person}{Zvonimir
  Rakamarić}.} \bibinfo{year}{2017}\natexlab{}.
\newblock \showarticletitle{Rigorous floating-point mixed-precision tuning}. In
  \bibinfo{booktitle}{\emph{Proceedings of the 44th {ACM} {SIGPLAN} {Symposium}
  on {Principles} of {Programming} {Languages}}} \emph{(\bibinfo{series}{{POPL}
  2017})}. \bibinfo{publisher}{Association for Computing Machinery},
  \bibinfo{address}{New York, NY, USA}, \bibinfo{pages}{300--315}.
\newblock
\showISBNx{978-1-4503-4660-3}
\urldef\tempurl%
\url{https://doi.org/10.1145/3009837.3009846}
\showDOI{\tempurl}


\bibitem[Chiang et~al\mbox{.}(2014)]%
        {s3fp}
\bibfield{author}{\bibinfo{person}{Wei-Fan Chiang}, \bibinfo{person}{Ganesh
  Gopalakrishnan}, \bibinfo{person}{Zvonimir Rakamaric}, {and}
  \bibinfo{person}{Alexey Solovyev}.} \bibinfo{year}{2014}\natexlab{}.
\newblock \showarticletitle{Efficient search for inputs causing high
  floating-point errors}. In \bibinfo{booktitle}{\emph{Proceedings of the 19th
  ACM SIGPLAN symposium on Principles and practice of parallel programming}}.
  \bibinfo{pages}{43--52}.
\newblock


\bibitem[Chowdhary et~al\mbox{.}(2020)]%
        {positdebug}
\bibfield{author}{\bibinfo{person}{Sangeeta Chowdhary}, \bibinfo{person}{Jay~P.
  Lim}, {and} \bibinfo{person}{Santosh Nagarakatte}.}
  \bibinfo{year}{2020}\natexlab{}.
\newblock \showarticletitle{Debugging and Detecting Numerical Errors in
  Computation with Posits}. In \bibinfo{booktitle}{\emph{Proceedings of the
  41st ACM SIGPLAN Conference on Programming Language Design and
  Implementation}} (London, UK) \emph{(\bibinfo{series}{PLDI 2020})}.
  \bibinfo{publisher}{Association for Computing Machinery},
  \bibinfo{address}{New York, NY, USA}, \bibinfo{pages}{731--746}.
\newblock
\showISBNx{9781450376136}
\urldef\tempurl%
\url{https://doi.org/10.1145/3385412.3386004}
\showDOI{\tempurl}


\bibitem[Cody and Waite(1980)]%
        {cody-waite}
\bibfield{author}{\bibinfo{person}{William~James Cody} {and}
  \bibinfo{person}{William Waite}.} \bibinfo{year}{1980}\natexlab{}.
\newblock \bibinfo{booktitle}{\emph{Software {Manual} for the {Elementary}
  {Functions} ({Prentice}-{Hall} series in computational mathematics)}}.
\newblock \bibinfo{publisher}{Prentice-Hall, Inc.}, \bibinfo{address}{USA}.
\newblock
\showISBNx{978-0-13-822064-8}


\bibitem[Damouche and Martel(2018)]%
        {salsa}
\bibfield{author}{\bibinfo{person}{Nasrine Damouche} {and}
  \bibinfo{person}{Matthieu Martel}.} \bibinfo{year}{2018}\natexlab{}.
\newblock \showarticletitle{Salsa: {An} {Automatic} {Tool} to {Improve} the
  {Numerical} {Accuracy} of {Programs}}. In \bibinfo{booktitle}{\emph{Kalpa
  {Publications} in {Computing}}}, Vol.~\bibinfo{volume}{5}.
  \bibinfo{publisher}{EasyChair}, \bibinfo{pages}{63--76}.
\newblock
\urldef\tempurl%
\url{https://doi.org/10.29007/j2fd}
\showDOI{\tempurl}
\newblock
\shownote{ISSN: 2515-1762}.


\bibitem[Daramy et~al\mbox{.}(2003)]%
        {crlibm}
\bibfield{author}{\bibinfo{person}{Catherine Daramy}, \bibinfo{person}{David
  Defour}, \bibinfo{person}{Florent Dinechin}, {and}
  \bibinfo{person}{Jean-Michel Muller}.} \bibinfo{year}{2003}\natexlab{}.
\newblock \showarticletitle{{CR}-{LIBM}: {A} correctly rounded elementary
  function library}.
\newblock \bibinfo{journal}{\emph{Proceedings of SPIE - The International
  Society for Optical Engineering}}  \bibinfo{volume}{5205}
  (\bibinfo{date}{Dec.} \bibinfo{year}{2003}).
\newblock
\urldef\tempurl%
\url{https://doi.org/10.1117/12.505591}
\showDOI{\tempurl}


\bibitem[Darulova et~al\mbox{.}(2018)]%
        {daisy}
\bibfield{author}{\bibinfo{person}{Eva Darulova}, \bibinfo{person}{Anastasiia
  Izycheva}, \bibinfo{person}{Fariha Nasir}, \bibinfo{person}{Fabian Ritter},
  \bibinfo{person}{Heiko Becker}, {and} \bibinfo{person}{Robert Bastian}.}
  \bibinfo{year}{2018}\natexlab{}.
\newblock \bibinfo{booktitle}{\emph{Daisy - Framework for Analysis and
  Optimization of Numerical Programs (Tool Paper)}}.
\newblock \bibinfo{pages}{270--287}.
\newblock
\showISBNx{978-3-319-89959-6}
\urldef\tempurl%
\url{https://doi.org/10.1007/978-3-319-89960-2_15}
\showDOI{\tempurl}


\bibitem[Darulova and Volkova(2019)]%
        {daisy-libm}
\bibfield{author}{\bibinfo{person}{Eva Darulova} {and}
  \bibinfo{person}{Anastasia Volkova}.} \bibinfo{year}{2019}\natexlab{}.
\newblock \showarticletitle{Sound Approximation of Programs with Elementary
  Functions}. In \bibinfo{booktitle}{\emph{Computer {Aided} {Verification}}}
  \emph{(\bibinfo{series}{Lecture {Notes} in {Computer} {Science}})},
  \bibfield{editor}{\bibinfo{person}{Isil Dillig} {and} \bibinfo{person}{Serdar
  Tasiran}} (Eds.). \bibinfo{publisher}{Springer International Publishing},
  \bibinfo{address}{Cham}, \bibinfo{pages}{174--183}.
\newblock
\showISBNx{978-3-030-25543-5}
\urldef\tempurl%
\url{https://doi.org/10.1007/978-3-030-25543-5_11}
\showDOI{\tempurl}


\bibitem[Das et~al\mbox{.}(2020)]%
        {satire}
\bibfield{author}{\bibinfo{person}{Arnab Das}, \bibinfo{person}{Ian Briggs},
  \bibinfo{person}{Ganesh Gopalakrishnan}, \bibinfo{person}{Sriram
  Krishnamoorthy}, {and} \bibinfo{person}{Pavel Panchekha}.}
  \bibinfo{year}{2020}\natexlab{}.
\newblock \showarticletitle{Scalable yet Rigorous Floating-Point Error
  Analysis}. In \bibinfo{booktitle}{\emph{Proceedings of the International
  Conference for High Performance Computing, Networking, Storage and Analysis}}
  (Atlanta, Georgia) \emph{(\bibinfo{series}{SC '20})}.
  \bibinfo{publisher}{IEEE Press}, Article \bibinfo{articleno}{51},
  \bibinfo{numpages}{14}~pages.
\newblock
\showISBNx{9781728199986}


\bibitem[de~Dinechin(2019)]%
        {flopoco}
\bibfield{author}{\bibinfo{person}{Florent de Dinechin}.}
  \bibinfo{year}{2019}\natexlab{}.
\newblock \showarticletitle{Reflections on 10 years of {FloPoCo}}. In
  \bibinfo{booktitle}{\emph{26th IEEE Symposium of Computer Arithmetic
  (ARITH-26)}}.
\newblock


\bibitem[Dekker(1971)]%
        {dekker}
\bibfield{author}{\bibinfo{person}{T.~J. Dekker}.}
  \bibinfo{year}{1971}\natexlab{}.
\newblock \showarticletitle{A floating-point technique for extending the
  available precision}.
\newblock \bibinfo{journal}{\emph{Numer. Math.}} \bibinfo{volume}{18},
  \bibinfo{number}{3} (\bibinfo{year}{1971}), \bibinfo{pages}{224--242}.
\newblock
\showISBNx{0945-3245}
\urldef\tempurl%
\url{https://doi.org/10.1007/BF01397083}
\showDOI{\tempurl}


\bibitem[Driscoll et~al\mbox{.}(2014)]%
        {chebfun}
\bibfield{author}{\bibinfo{person}{T.~A Driscoll}, \bibinfo{person}{N. Hale},
  {and} \bibinfo{person}{L.~N. Trefethen}.} \bibinfo{year}{2014}\natexlab{}.
\newblock \bibinfo{booktitle}{\emph{Chebfun Guide}}.
\newblock \bibinfo{publisher}{Pafnuty Publications}.
\newblock
\urldef\tempurl%
\url{http://www.chebfun.org/docs/guide/}
\showURL{%
\tempurl}


\bibitem[Fousse et~al\mbox{.}(2007)]%
        {mpfr}
\bibfield{author}{\bibinfo{person}{Laurent Fousse}, \bibinfo{person}{Guillaume
  Hanrot}, \bibinfo{person}{Vincent Lefèvre}, \bibinfo{person}{Patrick
  Pélissier}, {and} \bibinfo{person}{Paul Zimmermann}.}
  \bibinfo{year}{2007}\natexlab{}.
\newblock \showarticletitle{{MPFR}: {A} multiple-precision binary
  floating-point library with correct rounding}.
\newblock \bibinfo{journal}{\emph{ACM Trans. Math. Software}}
  \bibinfo{volume}{33}, \bibinfo{number}{2} (\bibinfo{date}{June}
  \bibinfo{year}{2007}), \bibinfo{pages}{13--es}.
\newblock
\showISSN{0098-3500}
\urldef\tempurl%
\url{https://doi.org/10.1145/1236463.1236468}
\showDOI{\tempurl}


\bibitem[{Free Software Foundation}(2020)]%
        {glibc-accuracy}
\bibfield{author}{\bibinfo{person}{{Free Software Foundation}}.}
  \bibinfo{year}{2020}\natexlab{}.
\newblock \bibinfo{title}{Errors in {Math} {Functions} ({The} {GNU} {C}
  {Library})}.
\newblock
\newblock
\urldef\tempurl%
\url{https://www.gnu.org/software/libc/manual/html_node/Errors-in-Math-Functions.html}
\showURL{%
\tempurl}


\bibitem[{FSF}(2020)]%
        {glibc}
\bibfield{author}{\bibinfo{person}{{FSF}}.} \bibinfo{year}{2020}\natexlab{}.
\newblock \bibinfo{title}{The GNU C Library}.
\newblock
\newblock
\urldef\tempurl%
\url{https://www.gnu.org/software/libc/manual/}
\showURL{%
\tempurl}


\bibitem[Group(2023)]%
        {llvmlibm}
\bibfield{author}{\bibinfo{person}{LLVM~Developer Group}.}
  \bibinfo{year}{2023}\natexlab{}.
\newblock \bibinfo{title}{Math Functions}.
\newblock
\newblock
\urldef\tempurl%
\url{https://libc.llvm.org/math/index.html}
\showURL{%
\tempurl}


\bibitem[Gulwani et~al\mbox{.}(2005)]%
        {egraph-intersect}
\bibfield{author}{\bibinfo{person}{Sumit Gulwani}, \bibinfo{person}{Ashish
  Tiwari}, {and} \bibinfo{person}{George~C. Necula}.}
  \bibinfo{year}{2005}\natexlab{}.
\newblock \showarticletitle{Join Algorithms for the Theory of Uninterpreted
  Functions}. In \bibinfo{booktitle}{\emph{FSTTCS 2004: Foundations of Software
  Technology and Theoretical Computer Science}},
  \bibfield{editor}{\bibinfo{person}{Kamal Lodaya} {and} \bibinfo{person}{Meena
  Mahajan}} (Eds.). \bibinfo{publisher}{Springer Berlin Heidelberg},
  \bibinfo{address}{Berlin, Heidelberg}, \bibinfo{pages}{311--323}.
\newblock
\showISBNx{978-3-540-30538-5}


\bibitem[Guo and Rubio-Gonz\'{a}lez(2020)]%
        {fpgen}
\bibfield{author}{\bibinfo{person}{Hui Guo} {and} \bibinfo{person}{Cindy
  Rubio-Gonz\'{a}lez}.} \bibinfo{year}{2020}\natexlab{}.
\newblock \showarticletitle{Efficient Generation of Error-Inducing
  Floating-Point Inputs via Symbolic Execution}. In
  \bibinfo{booktitle}{\emph{Proceedings of the ACM/IEEE 42nd International
  Conference on Software Engineering}} (Seoul, South Korea)
  \emph{(\bibinfo{series}{ICSE '20})}. \bibinfo{publisher}{Association for
  Computing Machinery}, \bibinfo{address}{New York, NY, USA},
  \bibinfo{pages}{1261--1272}.
\newblock
\showISBNx{9781450371216}
\urldef\tempurl%
\url{https://doi.org/10.1145/3377811.3380359}
\showDOI{\tempurl}


\bibitem[Guo and Rubio-González(2018)]%
        {hifptuner}
\bibfield{author}{\bibinfo{person}{Hui Guo} {and} \bibinfo{person}{Cindy
  Rubio-González}.} \bibinfo{year}{2018}\natexlab{}.
\newblock \showarticletitle{Exploiting community structure for floating-point
  precision tuning}. In \bibinfo{booktitle}{\emph{Proceedings of the 27th {ACM}
  {SIGSOFT} {International} {Symposium} on {Software} {Testing} and
  {Analysis}}} \emph{(\bibinfo{series}{{ISSTA} 2018})}.
  \bibinfo{publisher}{Association for Computing Machinery},
  \bibinfo{address}{New York, NY, USA}, \bibinfo{pages}{333--343}.
\newblock
\showISBNx{978-1-4503-5699-2}
\urldef\tempurl%
\url{https://doi.org/10.1145/3213846.3213862}
\showDOI{\tempurl}


\bibitem[Gustafson and Yonemoto(2017)]%
        {gappa}
\bibfield{author}{\bibinfo{person}{John~L. Gustafson} {and}
  \bibinfo{person}{Isaac~T. Yonemoto}.} \bibinfo{year}{2017}\natexlab{}.
\newblock \showarticletitle{Beating Floating Point at its Own Game: Posit
  Arithmetic}.
\newblock \bibinfo{journal}{\emph{Supercomput. Front. Innov.}}
  \bibinfo{volume}{4}, \bibinfo{number}{2} (\bibinfo{year}{2017}),
  \bibinfo{pages}{71--86}.
\newblock
\urldef\tempurl%
\url{https://doi.org/10.14529/jsfi170206}
\showDOI{\tempurl}


\bibitem[Hastings(1955)]%
        {cecil}
\bibfield{author}{\bibinfo{person}{Cecil Hastings}.}
  \bibinfo{year}{1955}\natexlab{}.
\newblock \showarticletitle{Approximations for Digital Computers}.
\newblock \bibinfo{journal}{\emph{Science}} \bibinfo{volume}{122},
  \bibinfo{number}{3170} (\bibinfo{year}{1955}), \bibinfo{pages}{602--602}.
\newblock
\urldef\tempurl%
\url{https://doi.org/10.1126/science.122.3170.602.a}
\showDOI{\tempurl}
\showeprint{https://www.science.org/doi/pdf/10.1126/science.122.3170.602.a}


\bibitem[Hida et~al\mbox{.}(2008)]%
        {dbailey}
\bibfield{author}{\bibinfo{person}{Yozo Hida}, \bibinfo{person}{Sherry Li},
  {and} \bibinfo{person}{David Bailey}.} \bibinfo{year}{2008}\natexlab{}.
\newblock \showarticletitle{Library for Double-Double and Quad-Double
  Arithmetic}.
\newblock  (\bibinfo{date}{01} \bibinfo{year}{2008}).
\newblock


\bibitem[Higham(2002)]%
        {higham}
\bibfield{author}{\bibinfo{person}{Nicholas~J. Higham}.}
  \bibinfo{year}{2002}\natexlab{}.
\newblock \bibinfo{booktitle}{\emph{Accuracy and {Stability} of {Numerical}
  {Algorithms}}}.
\newblock \bibinfo{publisher}{Society for Industrial and Applied Mathematics}.
\newblock
\showISBNx{978-0-89871-521-7}
\urldef\tempurl%
\url{https://doi.org/10.1137/1.9780898718027}
\showDOI{\tempurl}


\bibitem[Inc.(2022)]%
        {matlab}
\bibfield{author}{\bibinfo{person}{MathWorks Inc.}}
  \bibinfo{year}{2022}\natexlab{}.
\newblock \bibinfo{booktitle}{\emph{MATLAB version: 9.13.0 (R2022b)}}.
\newblock Natick, Massachusetts, United States.
\newblock
\urldef\tempurl%
\url{https://www.mathworks.com}
\showURL{%
\tempurl}


\bibitem[Intel(2020)]%
        {mkl}
\bibfield{author}{\bibinfo{person}{Intel}.} \bibinfo{year}{2020}\natexlab{}.
\newblock \bibinfo{title}{{Intel-Optimized} Math Library for Numerical
  Computing}.
\newblock
\newblock
\urldef\tempurl%
\url{http://software.intel.com/en-us/intel-mkl}
\showURL{%
\tempurl}


\bibitem[Johansson(2017)]%
        {arb}
\bibfield{author}{\bibinfo{person}{F. Johansson}.}
  \bibinfo{year}{2017}\natexlab{}.
\newblock \showarticletitle{Arb: efficient arbitrary-precision midpoint-radius
  interval arithmetic}.
\newblock \bibinfo{journal}{\emph{IEEE Trans. Comput.}}  \bibinfo{volume}{66}
  (\bibinfo{year}{2017}), \bibinfo{pages}{1281--1292}.
\newblock
Issue 8.
\urldef\tempurl%
\url{https://doi.org/10.1109/TC.2017.2690633}
\showDOI{\tempurl}


\bibitem[{Julia Math Project}(2021)]%
        {openlibm}
\bibfield{author}{\bibinfo{person}{{Julia Math Project}}.}
  \bibinfo{year}{2021}\natexlab{}.
\newblock \bibinfo{title}{{JuliaMath}/{OpenLibm}}.
\newblock
\newblock
\urldef\tempurl%
\url{https://github.com/JuliaMath/openlibm}
\showURL{%
\tempurl}


\bibitem[Kahan(1967)]%
        {kahan}
\bibfield{author}{\bibinfo{person}{William Kahan}.}
  \bibinfo{year}{1967}\natexlab{}.
\newblock \showarticletitle{7094-11 {System} support for numerical analysis}.
  In \bibinfo{booktitle}{\emph{Proceedings}}. \bibinfo{publisher}{US Army
  Research Office.}, \bibinfo{pages}{175}.
\newblock


\bibitem[Kahan(2004)]%
        {kahan-log}
\bibfield{author}{\bibinfo{person}{William Kahan}.}
  \bibinfo{year}{2004}\natexlab{}.
\newblock \bibinfo{title}{A {Logarithm} {Too} {Clever} by {Half}}.
\newblock
\newblock
\urldef\tempurl%
\url{http://people.eecs.berkeley.edu/~wkahan/LOG10HAF.TXT}
\showURL{%
\tempurl}


\bibitem[Kozen(1977)]%
        {egraphs}
\bibfield{author}{\bibinfo{person}{Dexter Kozen}.}
  \bibinfo{year}{1977}\natexlab{}.
\newblock \showarticletitle{Complexity of Finitely Presented Algebras}. In
  \bibinfo{booktitle}{\emph{Proceedings of the Ninth Annual ACM Symposium on
  Theory of Computing}} (Boulder, Colorado, USA) \emph{(\bibinfo{series}{STOC
  '77})}. \bibinfo{publisher}{Association for Computing Machinery},
  \bibinfo{address}{New York, NY, USA}, \bibinfo{pages}{164--177}.
\newblock
\showISBNx{9781450374095}
\urldef\tempurl%
\url{https://doi.org/10.1145/800105.803406}
\showDOI{\tempurl}


\bibitem[Kupriianova and Lauter(2014)]%
        {metalibm}
\bibfield{author}{\bibinfo{person}{Olga Kupriianova} {and}
  \bibinfo{person}{Christoph Lauter}.} \bibinfo{year}{2014}\natexlab{}.
\newblock \showarticletitle{Metalibm: {A} {Mathematical} {Functions} {Code}
  {Generator}}. In \bibinfo{booktitle}{\emph{Mathematical {Software} – {ICMS}
  2014}} \emph{(\bibinfo{series}{Lecture {Notes} in {Computer} {Science}})},
  \bibfield{editor}{\bibinfo{person}{Hoon Hong} {and} \bibinfo{person}{Chee
  Yap}} (Eds.). \bibinfo{publisher}{Springer}, \bibinfo{address}{Berlin,
  Heidelberg}, \bibinfo{pages}{713--717}.
\newblock
\showISBNx{978-3-662-44199-2}
\urldef\tempurl%
\url{https://doi.org/10.1007/978-3-662-44199-2_106}
\showDOI{\tempurl}


\bibitem[Lee et~al\mbox{.}(2017)]%
        {verified-math-h}
\bibfield{author}{\bibinfo{person}{Wonyeol Lee}, \bibinfo{person}{Rahul
  Sharma}, {and} \bibinfo{person}{Alex Aiken}.}
  \bibinfo{year}{2017}\natexlab{}.
\newblock \showarticletitle{On automatically proving the correctness of math.h
  implementations}.
\newblock \bibinfo{journal}{\emph{Proceedings of the ACM on Programming
  Languages}} \bibinfo{volume}{2}, \bibinfo{number}{POPL} (\bibinfo{date}{Dec.}
  \bibinfo{year}{2017}), \bibinfo{pages}{47:1--47:32}.
\newblock
\urldef\tempurl%
\url{https://doi.org/10.1145/3158135}
\showDOI{\tempurl}


\bibitem[Lim et~al\mbox{.}(2021)]%
        {jay-p-lim}
\bibfield{author}{\bibinfo{person}{Jay~P. Lim}, \bibinfo{person}{Mridul
  Aanjaneya}, \bibinfo{person}{John Gustafson}, {and} \bibinfo{person}{Santosh
  Nagarakatte}.} \bibinfo{year}{2021}\natexlab{}.
\newblock \showarticletitle{An Approach to Generate Correctly Rounded Math
  Libraries for New Floating Point Variants}.
\newblock \bibinfo{journal}{\emph{Proc. ACM Program. Lang.}}
  \bibinfo{volume}{5}, \bibinfo{number}{POPL}, Article \bibinfo{articleno}{29}
  (\bibinfo{date}{jan} \bibinfo{year}{2021}), \bibinfo{numpages}{30}~pages.
\newblock
\urldef\tempurl%
\url{https://doi.org/10.1145/3434310}
\showDOI{\tempurl}


\bibitem[Lim and Nagarakatte(2022)]%
        {rlibm1}
\bibfield{author}{\bibinfo{person}{Jay~P. Lim} {and} \bibinfo{person}{Santosh
  Nagarakatte}.} \bibinfo{year}{2022}\natexlab{}.
\newblock \showarticletitle{One Polynomial Approximation to Produce Correctly
  Rounded Results of an Elementary Function for Multiple Representations and
  Rounding Modes}.
\newblock \bibinfo{journal}{\emph{Proc. ACM Program. Lang.}}
  \bibinfo{volume}{6}, \bibinfo{number}{POPL}, Article \bibinfo{articleno}{3}
  (\bibinfo{date}{jan} \bibinfo{year}{2022}), \bibinfo{numpages}{28}~pages.
\newblock
\urldef\tempurl%
\url{https://doi.org/10.1145/3498664}
\showDOI{\tempurl}


\bibitem[{Maplesoft, a division of Waterloo Maple Inc..}(2019)]%
        {maple}
\bibfield{author}{\bibinfo{person}{{Maplesoft, a division of Waterloo Maple
  Inc..}}} \bibinfo{year}{2019}\natexlab{}.
\newblock \bibinfo{booktitle}{\emph{Maple}}.
\newblock Waterloo, Ontario.
\newblock
\urldef\tempurl%
\url{https://www.maplesoft.com}
\showURL{%
\tempurl}


\bibitem[Micikevicius et~al\mbox{.}(2022)]%
        {fp8}
\bibfield{author}{\bibinfo{person}{Paulius Micikevicius},
  \bibinfo{person}{Dusan Stosic}, \bibinfo{person}{Neil Burgess},
  \bibinfo{person}{Marius Cornea}, \bibinfo{person}{Pradeep Dubey},
  \bibinfo{person}{Richard Grisenthwaite}, \bibinfo{person}{Sangwon Ha},
  \bibinfo{person}{Alexander Heinecke}, \bibinfo{person}{Patrick Judd},
  \bibinfo{person}{John Kamalu}, \bibinfo{person}{Naveen Mellempudi},
  \bibinfo{person}{Stuart Oberman}, \bibinfo{person}{Mohammad Shoeybi},
  \bibinfo{person}{Michael Siu}, {and} \bibinfo{person}{Hao Wu}.}
  \bibinfo{year}{2022}\natexlab{}.
\newblock \bibinfo{title}{FP8 Formats for Deep Learning}.
\newblock
\newblock
\showeprint[arxiv]{2209.05433}~[cs.LG]


\bibitem[Microsystems(1993)]%
        {fdlibm}
\bibfield{author}{\bibinfo{person}{Sun Microsystems}.}
  \bibinfo{year}{1993}\natexlab{}.
\newblock \showarticletitle{A Freely Distributable Libm}.
\newblock  (\bibinfo{year}{1993}).
\newblock
\urldef\tempurl%
\url{https://www.netlib.org/fdlibm/}
\showURL{%
\tempurl}


\bibitem[Moshier(1992)]%
        {cephes}
\bibfield{author}{\bibinfo{person}{Stephen~L Moshier}.}
  \bibinfo{year}{1992}\natexlab{}.
\newblock \bibinfo{title}{Cephes mathematical library}.
\newblock
\newblock
\urldef\tempurl%
\url{http://www.netlib.org/cephes/}
\showURL{%
\tempurl}


\bibitem[Muller(2016)]%
        {muller}
\bibfield{author}{\bibinfo{person}{Jean-Michel Muller}.}
  \bibinfo{year}{2016}\natexlab{}.
\newblock \bibinfo{booktitle}{\emph{Elementary {Functions}: {Algorithms} and
  {Implementation}} (\bibinfo{edition}{3} ed.)}.
\newblock \bibinfo{publisher}{Birkhäuser Basel}.
\newblock
\showISBNx{978-1-4899-7981-0}
\urldef\tempurl%
\url{https://doi.org/10.1007/978-1-4899-7983-4}
\showDOI{\tempurl}


\bibitem[Panchekha et~al\mbox{.}(2015)]%
        {herbie}
\bibfield{author}{\bibinfo{person}{Pavel Panchekha}, \bibinfo{person}{Alex
  Sanchez-Stern}, \bibinfo{person}{James~R. Wilcox}, {and}
  \bibinfo{person}{Zachary Tatlock}.} \bibinfo{year}{2015}\natexlab{}.
\newblock \showarticletitle{Automatically Improving Accuracy for Floating Point
  Expressions}. In \bibinfo{booktitle}{\emph{Proceedings of the 36th ACM
  SIGPLAN Conference on Programming Language Design and Implementation}}
  (Portland, OR, USA) \emph{(\bibinfo{series}{PLDI '15})}.
  \bibinfo{publisher}{Association for Computing Machinery},
  \bibinfo{address}{New York, NY, USA}, \bibinfo{pages}{1--11}.
\newblock
\showISBNx{9781450334686}
\urldef\tempurl%
\url{https://doi.org/10.1145/2737924.2737959}
\showDOI{\tempurl}


\bibitem[Piparo et~al\mbox{.}(2014)]%
        {vdt}
\bibfield{author}{\bibinfo{person}{Danilo Piparo}, \bibinfo{person}{Vincenzo
  Innocente}, {and} \bibinfo{person}{Thomas Hauth}.}
  \bibinfo{year}{2014}\natexlab{}.
\newblock \showarticletitle{Speeding up {HEP} experiment software with a
  library of fast and auto-vectorisable mathematical functions}.
\newblock \bibinfo{journal}{\emph{Journal of Physics: Conference Series}}
  \bibinfo{volume}{513}, \bibinfo{number}{5} (\bibinfo{date}{June}
  \bibinfo{year}{2014}), \bibinfo{pages}{052027}.
\newblock
\showISSN{1742-6596}
\urldef\tempurl%
\url{https://doi.org/10.1088/1742-6596/513/5/052027}
\showDOI{\tempurl}
\newblock
\shownote{Publisher: IOP Publishing}.


\bibitem[Remez and Gavrilyuk(1968)]%
        {remez}
\bibfield{author}{\bibinfo{person}{Evgenii~Yakovlevich Remez} {and}
  \bibinfo{person}{Vera~Timofeevna Gavrilyuk}.}
  \bibinfo{year}{1968}\natexlab{}.
\newblock \showarticletitle{A theorem on interpolation functions which provides
  a fundamental approach to the treatment of general analogs of the method of
  successive Chebyshev interpolations}. In \bibinfo{booktitle}{\emph{Doklady
  Akademii Nauk}}, Vol.~\bibinfo{volume}{183}. Russian Academy of Sciences,
  \bibinfo{pages}{750--753}.
\newblock


\bibitem[Research~Inc.(2023)]%
        {mathematica}
\bibfield{author}{\bibinfo{person}{Wolfram Research~Inc.}}
  \bibinfo{year}{2023}\natexlab{}.
\newblock \bibinfo{title}{Mathematica, {V}ersion 13.3}.
\newblock
\newblock
\urldef\tempurl%
\url{https://www.wolfram.com/mathematica}
\showURL{%
\tempurl}
\newblock
\shownote{Champaign, IL}.


\bibitem[Rubio-González et~al\mbox{.}(2013)]%
        {precimonious}
\bibfield{author}{\bibinfo{person}{C. Rubio-González}, \bibinfo{person}{{Cuong
  Nguyen}}, \bibinfo{person}{{Hong Diep Nguyen}}, \bibinfo{person}{J. Demmel},
  \bibinfo{person}{W. Kahan}, \bibinfo{person}{K. Sen}, \bibinfo{person}{D.~H.
  Bailey}, \bibinfo{person}{C. Iancu}, {and} \bibinfo{person}{D. Hough}.}
  \bibinfo{year}{2013}\natexlab{}.
\newblock \showarticletitle{{Precimonious}: Tuning assistant for floating-point
  precision}. In \bibinfo{booktitle}{\emph{{SC} '13: {Proceedings} of the
  {International} {Conference} on {High} {Performance} {Computing},
  {Networking}, {Storage} and {Analysis}}}. \bibinfo{pages}{1--12}.
\newblock
\urldef\tempurl%
\url{https://doi.org/10.1145/2503210.2503296}
\showDOI{\tempurl}
\newblock
\shownote{ISSN: 2167-4337}.


\bibitem[S. et~al\mbox{.}(2010)]%
        {sollya}
\bibfield{author}{\bibinfo{person}{Chevillard S.}, \bibinfo{person}{Joldeş
  M.}, {and} \bibinfo{person}{Lauter C.}} \bibinfo{year}{2010}\natexlab{}.
\newblock \showarticletitle{Sollya: {An} {Environment} for the {Development} of
  {Numerical} {Codes}}. In \bibinfo{booktitle}{\emph{Mathematical {Software} -
  {ICMS} 2010}} \emph{(\bibinfo{series}{Lecture {Notes} in {Computer}
  {Science}}, Vol.~\bibinfo{volume}{6327})},
  \bibfield{editor}{\bibinfo{person}{{K. Fukuda}}, \bibinfo{person}{{J. van der
  Hoeven}}, \bibinfo{person}{{M. Joswig}}, {and} \bibinfo{person}{{N.
  Takayama}}} (Eds.). \bibinfo{publisher}{Springer},
  \bibinfo{address}{Heidelberg, Germany}, \bibinfo{pages}{28--31}.
\newblock


\bibitem[Saiki et~al\mbox{.}(2021)]%
        {pherbie}
\bibfield{author}{\bibinfo{person}{Brett Saiki}, \bibinfo{person}{Oliver
  Flatt}, \bibinfo{person}{Chandrakana Nandi}, \bibinfo{person}{Pavel
  Panchekha}, {and} \bibinfo{person}{Zachary Tatlock}.}
  \bibinfo{year}{2021}\natexlab{}.
\newblock \showarticletitle{Combining Precision Tuning and Rewriting}. In
  \bibinfo{booktitle}{\emph{2021 IEEE 28th Symposium on Computer Arithmetic
  (ARITH)}}.
\newblock


\bibitem[Sanchez-Stern et~al\mbox{.}(2018)]%
        {herbgrind}
\bibfield{author}{\bibinfo{person}{Alex Sanchez-Stern}, \bibinfo{person}{Pavel
  Panchekha}, \bibinfo{person}{Sorin Lerner}, {and} \bibinfo{person}{Zachary
  Tatlock}.} \bibinfo{year}{2018}\natexlab{}.
\newblock \showarticletitle{Finding Root Causes of Floating Point Error}.
\newblock \bibinfo{journal}{\emph{SIGPLAN Not.}} \bibinfo{volume}{53},
  \bibinfo{number}{4} (\bibinfo{date}{jun} \bibinfo{year}{2018}),
  \bibinfo{pages}{256--269}.
\newblock
\showISSN{0362-1340}
\urldef\tempurl%
\url{https://doi.org/10.1145/3296979.3192411}
\showDOI{\tempurl}


\bibitem[Sibidanov et~al\mbox{.}(2022)]%
        {core-math}
\bibfield{author}{\bibinfo{person}{Alexei Sibidanov}, \bibinfo{person}{Paul
  Zimmermann}, {and} \bibinfo{person}{Stéphane Glondu}.}
  \bibinfo{year}{2022}\natexlab{}.
\newblock \showarticletitle{The CORE-MATH Project}. In
  \bibinfo{booktitle}{\emph{2022 IEEE 29th Symposium on Computer Arithmetic
  (ARITH)}}. \bibinfo{pages}{26--34}.
\newblock
\urldef\tempurl%
\url{https://doi.org/10.1109/ARITH54963.2022.00014}
\showDOI{\tempurl}


\bibitem[Solovyev et~al\mbox{.}(2018)]%
        {fptaylor}
\bibfield{author}{\bibinfo{person}{Alexey Solovyev}, \bibinfo{person}{Marek~S.
  Baranowski}, \bibinfo{person}{Ian Briggs}, \bibinfo{person}{Charles
  Jacobsen}, \bibinfo{person}{Zvonimir Rakamari\'{c}}, {and}
  \bibinfo{person}{Ganesh Gopalakrishnan}.} \bibinfo{year}{2018}\natexlab{}.
\newblock \showarticletitle{Rigorous Estimation of Floating-Point Round-Off
  Errors with Symbolic Taylor Expansions}.
\newblock \bibinfo{journal}{\emph{ACM Trans. Program. Lang. Syst.}}
  \bibinfo{volume}{41}, \bibinfo{number}{1}, Article \bibinfo{articleno}{2}
  (\bibinfo{date}{dec} \bibinfo{year}{2018}), \bibinfo{numpages}{39}~pages.
\newblock
\showISSN{0164-0925}
\urldef\tempurl%
\url{https://doi.org/10.1145/3230733}
\showDOI{\tempurl}


\bibitem[Sterbenz(1974)]%
        {sterbenz}
\bibfield{author}{\bibinfo{person}{Pat~H Sterbenz}.}
  \bibinfo{year}{1974}\natexlab{}.
\newblock \bibinfo{booktitle}{\emph{{Floating-point computation}}}.
\newblock \bibinfo{publisher}{Prentice-Hall}, \bibinfo{address}{Englewood
  Cliffs, NJ}.
\newblock
\urldef\tempurl%
\url{https://cds.cern.ch/record/268412}
\showURL{%
\tempurl}


\bibitem[Taylor(1717)]%
        {taylor}
\bibfield{author}{\bibinfo{person}{Brook Taylor}.}
  \bibinfo{year}{1717}\natexlab{}.
\newblock \bibinfo{booktitle}{\emph{Methodus incrementorum directa \&
  inversa}}.
\newblock \bibinfo{publisher}{Inny}.
\newblock


\bibitem[Tch{\'e}bychev(1858)]%
        {chebyshev}
\bibfield{author}{\bibinfo{person}{P Tch{\'e}bychev}.}
  \bibinfo{year}{1858}\natexlab{}.
\newblock \bibinfo{booktitle}{\emph{Sur les questions de minima qui se
  rattechent a la rapr{\'e}sentation aproximative des fonctions}}.
\newblock \bibinfo{publisher}{Imprimerie de l'Academie Imp{\'e}riale des
  Sciences}.
\newblock


\bibitem[{The Sage Developers}(2023)]%
        {sagemath}
\bibfield{author}{\bibinfo{person}{{The Sage Developers}}.}
  \bibinfo{year}{2023}\natexlab{}.
\newblock \bibinfo{booktitle}{\emph{{S}ageMath, the {S}age {M}athematics
  {S}oftware {S}ystem ({V}ersion x.y.z)}}.
\newblock
\newblock
\shownote{{\tt https://www.sagemath.org}}.


\bibitem[Wang and Kanwar(2019)]%
        {bfloat16}
\bibfield{author}{\bibinfo{person}{Shibo Wang} {and} \bibinfo{person}{Pankaj
  Kanwar}.} \bibinfo{year}{2019}\natexlab{}.
\newblock \bibinfo{title}{{BFloat16}: {The} secret to high performance on
  {Cloud} {TPUs}}.
\newblock
\newblock
\urldef\tempurl%
\url{https://cloud.google.com/blog/products/ai-machine-learning/bfloat16-the-secret-to-high-performance-on-cloud-tpus/}
\showURL{%
\tempurl}


\bibitem[Willsey et~al\mbox{.}(2021)]%
        {egg}
\bibfield{author}{\bibinfo{person}{Max Willsey}, \bibinfo{person}{Chandrakana
  Nandi}, \bibinfo{person}{Yisu~Remy Wang}, \bibinfo{person}{Oliver Flatt},
  \bibinfo{person}{Zachary Tatlock}, {and} \bibinfo{person}{Pavel Panchekha}.}
  \bibinfo{year}{2021}\natexlab{}.
\newblock \showarticletitle{Egg: Fast and Extensible Equality Saturation}.
\newblock \bibinfo{journal}{\emph{Proc. ACM Program. Lang.}}
  \bibinfo{volume}{5}, \bibinfo{number}{POPL}, Article \bibinfo{articleno}{23}
  (\bibinfo{date}{jan} \bibinfo{year}{2021}), \bibinfo{numpages}{29}~pages.
\newblock
\urldef\tempurl%
\url{https://doi.org/10.1145/3434304}
\showDOI{\tempurl}


\bibitem[Zimmermann(2021)]%
        {zimmermann}
\bibfield{author}{\bibinfo{person}{Paul Zimmermann}.}
  \bibinfo{year}{2021}\natexlab{}.
\newblock \bibinfo{title}{{Accuracy of Mathematical Functions in Single,
  Double, Extended Double and Quadruple Precision}}.  (\bibinfo{date}{Feb.}
  \bibinfo{year}{2021}).
\newblock
\urldef\tempurl%
\url{https://hal.inria.fr/hal-03141101}
\showURL{%
\tempurl}
\newblock
\shownote{working paper or preprint}.


\end{thebibliography}

\end{document}